\documentstyle[12pt]{article}

\catcode`\@=11
\long\def\@makefntext#1{
\protect\noindent \hbox to 3.2pt {\hskip-.9pt  
$^{{\ninerm\@thefnmark}}$\hfil}#1\hfill}		

\def\@makefnmark{\hbox to 0pt{$^{\@thefnmark}$\hss}}  
	
\def\ps@myheadings{\let\@mkboth\@gobbletwo
\def\@oddhead{\hbox{}
\rightmark\hfil\ninerm\thepage}   
\def\@oddfoot{}\def\@evenhead{\ninerm\thepage\hfil
\leftmark\hbox{}}\def\@evenfoot{}
\def\sectionmark##1{}\def\subsectionmark##1{}}

\renewcommand{\thefootnote}{\fnsymbol{footnote}}

\newcounter{sectionc}\newcounter{subsectionc}\newcounter{subsubsectionc}
\renewcommand{\section}[1] {\vspace*{0.6cm}\addtocounter{sectionc}{1} 
\setcounter{subsectionc}{0}\setcounter{subsubsectionc}{0}\noindent 
	{\normalsize\bf\thesectionc. #1}\par\vspace*{0.4cm}}
\renewcommand{\subsection}[1] {\vspace*{0.6cm}\addtocounter{subsectionc}{1} 
	\setcounter{subsubsectionc}{0}\noindent 
	{\normalsize\it\thesectionc.\thesubsectionc. #1}\par\vspace*{0.4cm}}
\renewcommand{\subsubsection}[1]
{\vspace*{0.6cm}\addtocounter{subsubsectionc}{1}
	\noindent {\normalsize\rm\thesectionc.\thesubsectionc.\thesubsubsectionc. 
	#1}\par\vspace*{0.4cm}}

\newcounter{appendixc}
\newcounter{subappendixc}[appendixc]
\newcounter{subsubappendixc}[subappendixc]

\renewcommand{\appendix}[1] {\vspace*{0.6cm}
        \refstepcounter{appendixc}
        \setcounter{figure}{0}
        \setcounter{table}{0}
        \setcounter{equation}{0}
        \renewcommand{\thefigure}{\Alph{appendixc}.\arabic{figure}}
        \renewcommand{\thetable}{\Alph{appendixc}.\arabic{table}}
        \renewcommand{\theappendixc}{\Alph{appendixc}}
        \renewcommand{\theequation}{\Alph{appendixc}.\arabic{equation}}
        \noindent{\bf Appendix \theappendixc #1}\par\vspace*{0.4cm}}

\def\abstracts#1{{
	\centering{\begin{minipage}{12.2truecm}\footnotesize\baselineskip=12pt\noindent
	\centerline{\footnotesize ABSTRACT}\vspace*{0.3cm}
	\parindent=0pt #1
	\end{minipage}}\par}} 

\renewenvironment{thebibliography}[1]
	{\begin{list}{\arabic{enumi}.}
	{\usecounter{enumi}\setlength{\parsep}{0pt}
\setlength{\leftmargin 1.25cm}{\rightmargin 0pt}
	 \setlength{\itemsep}{0pt} \settowidth
	{\labelwidth}{#1.}\sloppy}}{\end{list}}

\newcounter{itemlistc}
\newcounter{romanlistc}
\newcounter{alphlistc}
\newcounter{arabiclistc}

\newcommand{\fcaption}[1]{
        \refstepcounter{figure}
        \setbox\@tempboxa = \hbox{\footnotesize Fig.~\thefigure. #1}
        \ifdim \wd\@tempboxa > 6in
           {\begin{center}
        \parbox{6in}{\footnotesize\baselineskip=12pt Fig.~\thefigure. #1}
            \end{center}}
        \else
             {\begin{center}
             {\footnotesize Fig.~\thefigure. #1}
              \end{center}}
        \fi}
\newcommand{\ffcaption}[1]{
        \refstepcounter{figure}
        \setbox\@tempboxa = \hbox{\footnotesize Fig.~\thefigure. #1}
        \ifdim \wd\@tempboxa > 6in
           {\begin{center}
        \parbox{6in}{\footnotesize\baselineskip=12pt Fig.~\thefigure. #1}
            \end{center}}
        \else
             {\begin{center}
             {\footnotesize Fig.~\thefigure. #1}
              \end{center}}
        \fi}

\newcommand{\tcaption}[1]{
        \refstepcounter{table}
        \setbox\@tempboxa = \hbox{\footnotesize Table~\thetable. #1}
        \ifdim \wd\@tempboxa > 6in
           {\begin{center}
        \parbox{6in}{\footnotesize\baselineskip=12pt Table~\thetable. #1}
            \end{center}}
        \else
             {\begin{center}
             {\footnotesize Table~\thetable. #1}
              \end{center}}
        \fi}

\def\@citex[#1]#2{\if@filesw\immediate\write\@auxout
	{\string\citation{#2}}\fi
\def\@citea{}\@cite{\@for\@citeb:=#2\do
	{\@citea\def\@citea{,}\@ifundefined
	{b@\@citeb}{{\bf ?}\@warning
	{Citation `\@citeb' on page \thepage \space undefined}}
	{\csname b@\@citeb\endcsname}}}{#1}}

\newif\if@cghi
\def\cite{\@cghitrue\@ifnextchar [{\@tempswatrue
	\@citex}{\@tempswafalse\@citex[]}}
\def\citelow{\@cghifalse\@ifnextchar [{\@tempswatrue
	\@citex}{\@tempswafalse\@citex[]}}
\def\@cite#1#2{{$\null^{#1}$\if@tempswa\typeout
	{IJCGA warning: optional citation argument 
	ignored: `#2'} \fi}}

 1
\font\twelverm=cmr10  scaled\magstep 1
 1

\font\ninerm=cmr9

\def\spose#1{\hbox to 0pt{#1\hss}}
\def\lsim{\mathrel{\spose{\lower 3pt\hbox{$\mathchar"218$}}
     \raise 2.0pt\hbox{$\mathchar"13C$}}}
\def\gsim{\mathrel{\spose{\lower 3pt\hbox{$\mathchar"218$}}
     \raise 2.0pt\hbox{$\mathchar"13E$}}}
\def\simpropto{\mathrel{\spose{\lower 3pt\hbox{$\mathchar"218$}}
     \raise 2.0pt\hbox{$\propto$}}}
\newcommand{\beqn}{\begin{eqnarray}}
\newcommand{\eeqn}{\end{eqnarray}}

\def\ibid{{\sl ibid.}}

\def\hc{\rm H.c.}
\def\ie{ {\it i.e.}}
\def\eg{ {\it e.g.}}
\def\etal{ {\it et al.}}

\def\AJ#1#2#3{{\sl Astr. J.} {\bf #1}, #2 (#3)}
\def\ZPC#1#2#3{{\sl Z.~Phys.} {\bf C#1}, #2 (#3)}

\def\PRL#1#2#3{{\sl Phys. Rev. Lett.} {\bf #1}, #2 (#3)}
\def\PRD#1#2#3{{\sl Phys. Rev.} {\bf D#1}, #2 (#3)}
\def\PLB#1#2#3{{\sl Phys. Lett.} {\bf B#1}, #2 (#3)}
\def\PREP#1#2#3{{\sl Phys. Rep.} {\bf #1}, #2 (#3)}
\def\NPB#1#2#3{{\sl Nucl. Phys.} {\bf B#1}, #2 (#3)}
\def\citenum#1{{\def\@cite##1##2{##1}\cite{#1}}}

\def\gev{{\rm GeV }}

\def\ev{{\rm eV }}

\def\sec{{\rm sec}}

\def\mgut{M_{\rm GUT}}
\def\mpl{M_{\rm P}}
\def\mz{m_{\rm z}}

\def\vev#1{{\langle#1\rangle}}

\def\half{{1\over 2}}

\def\cosb{\cos \beta}
\def\tanb{\tan \beta}
\def\barv{\overline{v}}

\begin{document}

\centerline{\normalsize\bf SUSY WITHOUT $R$-PARITY:}
\baselineskip=22pt
\centerline{\normalsize\bf SYMMETRY BREAKING AND LSP-PHENOMENOLOGY}
\baselineskip=16pt

\centerline{\footnotesize Ralf Hempfling}
\baselineskip=13pt
\centerline{\footnotesize\it  Davis Institute for High Energy Physics}
\baselineskip=12pt
\centerline{\footnotesize\it Davis, CA 95616, USA}
\centerline{\footnotesize E-mail: hempf@bethe.ucdavis.edu}
\vspace*{0.3cm}

\vspace*{0.9cm}
\abstracts{
We have studied the predictions for the
LSP decay within the framework of a radiatively broken
unified supergravity model without $R$-parity.
Assuming that Higgs/slepton mixing is the only source of
$R$-parity breaking and responsible for the observed neutrino
oscillations we obtain predictons for the LSP
life-time and branching fractions.
}
 
\normalsize\baselineskip=15pt
\setcounter{footnote}{0}
\renewcommand{\thefootnote}{\alph{footnote}}
\section{Motivation}
Supersymmetry\cite{susyreview} is presently the most popular
attempt to solve the hierarchy problem of the
standard model (SM).
Here, the cancellation of quadratic divergences
is guaranteed and, hence, any mass scale is stable under radiative
corrections. 
The most economical candidate for a realistic model
is the minimal supersymmetric extension of the SM (MSSM).
In the SM baryon and lepton number are protected by
an accidental symmetry (\ie there is no gauge and Lorentz
invariant term of dimension 4 or less that violates
$B$ or $L$ via perturbative effects).
This no longer holds in the MSSM due to the existence of superpartners.
One way to assure Baryon and Lepton number conservation
(and hence the stability of the proton)
is to impose by hand 
a discrete, multiplicative symmetry called
$R$-parity\cite{r-parity},
$R_p = (-1)^{2S+3B+L}$,
where $S$, $B$ and $L$ are the spin, baryon and lepton numbers,
respectively.
Aside from the long proton life-time,
$R_p$ conserving models have the very attractive feature
that the lightest supersymmetric particle (LSP)
is stable and a good cold dark matter candidate\cite{cdm}.
On the other hand, there is strong experimental evidence for
neutrino oscillations\cite{solarn,atmosphericn,lsnd}
which can be accounted for
if $R_p$ is broken\cite{npb478}

In this paper, we will investigate
the LSP life-time in a SUSY-GUT
scenario where $R_p$ is broken explicitly via
dimension 2
terms\cite{npb478,suzuki,yossi,muk}.
We have discussed this model  detail in ref.~\citenum{npb478}
where the emphasis was on
neutrino phenomenology in the frame-work
of radiative electro-weak spontaneous symmetry breaking
(REWSSB)\cite{radssb}.
Here, we are particularly interested in the implications for high energy
collider phenomenology.
We will focus attention of the case that the LSP
is a neutralino.
This is the most interesting case, since it 
occurs naturally over most of the SUSY parameter space.
Note, however, that
in models with broken $R_p$ there is no
theoretical/cosmological prejudice concerning the color or
electric charge of
the LSP. 
The only requirement is that the LSP life-time
is sufficiently short ($\tau_{LSP}\lsim 1~\sec$)
so as not to disturb big bang nucleosynthesis\cite{subir}
or sufficiently long
[$\tau_{LSP}\gsim 10^{24}~\sec/ B(LSP\rightarrow X \nu_e)$]
so as not to lead to an unacceptable
distortion of the cosmic microwave background\cite{diwan}.

The most general  gauge invariant superpotential
can be written as
\beqn
W &=& \half y^L_{I J k} \hat L_I \hat L_J \hat E^c_k
   +      y^D_{I j k} \hat L_I \hat Q_j \hat D^c_k
   -      y^U_{  j k} \hat H   \hat Q_j \hat U^c_k
   -      \mu_I   \hat L_I \hat H
   + \half \bar y^D_{i j k}\hat  D^c_i\hat  D^c_j\hat  U^c_k\,,
\label{defw}
\eeqn
where the supermultiplets are denoted by a hat.
The left-handed lepton supermultiplets
are denoted by $\hat L_i$ ($i = 1,2,3$)
and the Higgs superfield coupling to the down-type quarks
is denoted by $\hat L_0$.

Let us first determine the meaning of the various terms of
eq.~\ref{defw}. Here, $y_{0jk}^L$, $y_{0jk}^D$ and $y_{jk}^U$
denote the lepton, down-type and up-type Yukawa couplings, respectively,
and $\mu_0$ is the Higgs mass parameter.
However, in contrast to the SM the MSSM allows for 
renormalizable baryon [lepton] number violating interactions
$\bar y_{ijk}^D$ [$y_{ijk}^L$ and $\mu_i$].
These couplings are constrained from above by experiment.
The most model independent constraints can be obtained from 
collider experiments\cite{collider-c} or neutrino
physics \cite{neutrino-c1}.
It turns out that the individual
lepton and baryon number
violating couplings only have to be smaller
than $O(10^{-3}\sim {\rm few}\times 10^{-1})$
with the exception of
$\bar y_{121}^D\lsim 10^{-7}$
from heavy nuclei decay\cite{sher}.
Thus, the $R_p$ violating couplings need not
be much more suppressed than the
lepton and baryon number preserving Yukawa couplings.
(remember that \eg, $y^D_{011} \simeq 3\times 10^{-5}/\cosb$).
Somewhat stronger but more model dependent
constraints can be derived
from cosmology\cite{cosmology-c}.

However, the experimental exclusion area can be strongly enhanced by
imposing theoretical constraints:
in the minimal SU(5) SUSY-GUT model, the right-handed leptons,
the right-handed up-type quarks and the left-handed quarks are embedded
in a 10-dim representation,
${\bf 10}_i = E^c_i\oplus U^c_i \oplus Q_i$.
The right-handed down-type 
quarks and the left-handed leptons are embedded
in a 5-dim representation,
${\bf \overline{5}}_i = D_i\oplus L_i$.
The two Higgs doublets are embedded together
with two proton decay mediating
colored triplets, $T$ and $D_0$, in 5-dim representations,
${\bf \overline{5}}_0 = D_0\oplus L_0$ and
${\bf           5 }   = T\oplus H$.
Hence, both the lepton and the baryon number violating 
interactions arise from the term
\beqn
W_{\rm GUT} = \half y_{i j k} {\bf \overline{5}}_i 
{\bf \overline{5}}_j {\bf 10}_k
\,,\label{wsu5}
\eeqn
where the boundary conditions at the GUT-scale, $\mgut$, are given by
$y^L_{i j k} = y^D_{i k j} = \bar y_{i j k}^D = y_{i j k}$.
These relations, which predict the down-type quark masses
correctly to within a factor of 3, are expected to also hold
at a comparable level
for the $R_p$ violating couplings.
Thus, in general the baryon and lepton number
violating couplings are correlated in SUSY GUT models.
This leads to very strong constraints
on any $y_{i j k}$ from proton
stability which are much stronger than any
constraint on individual Yukawa couplings\cite{smirnov}.
However, it does not place any constraints on the coefficients of
of dimension 2 terms, $\mu_i$,
which are the subject of this paper.

\begin{figure}
\vspace*{13pt}
\vspace*{2.0truein}      
\includegraphics{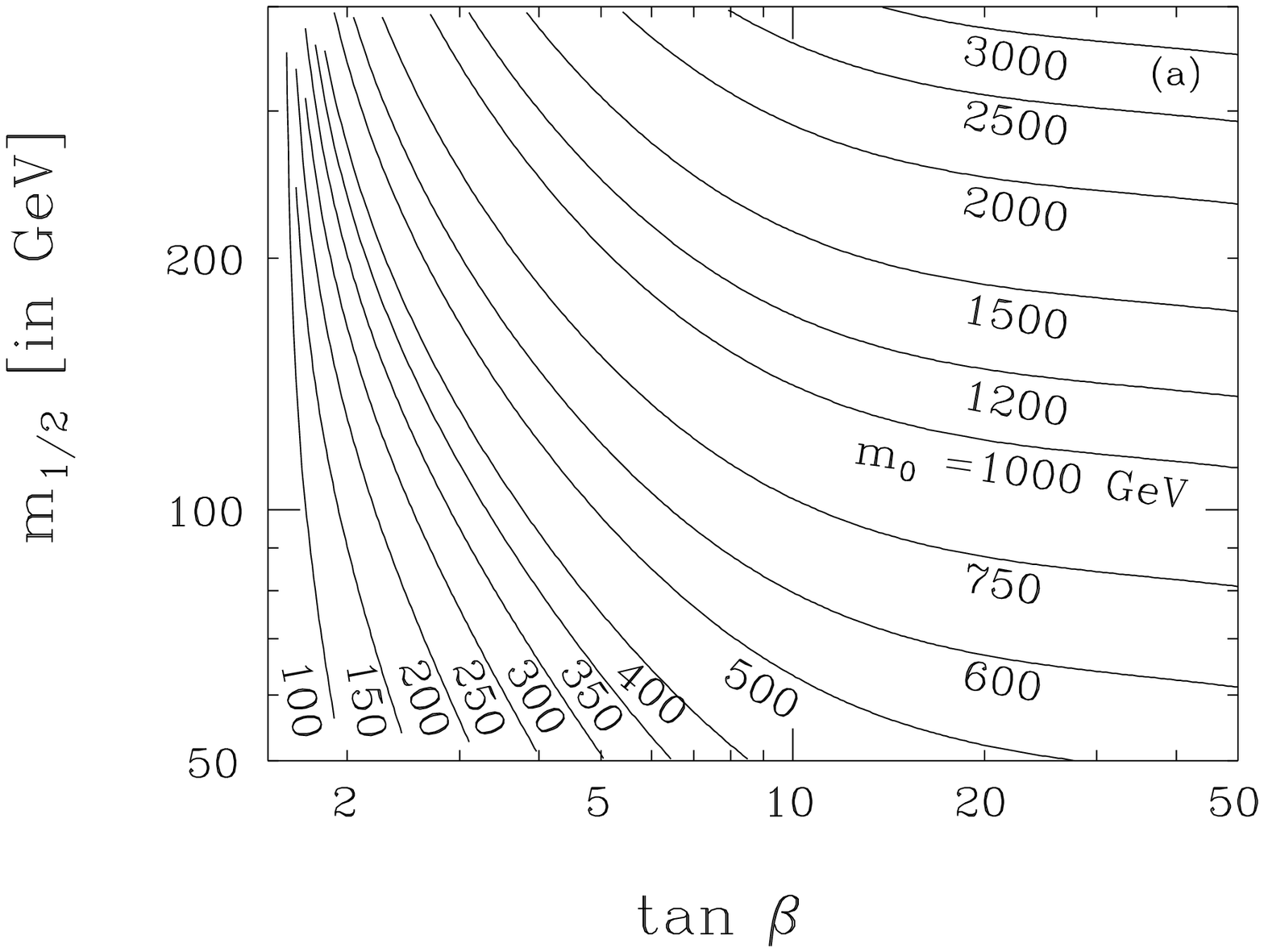}
\includegraphics{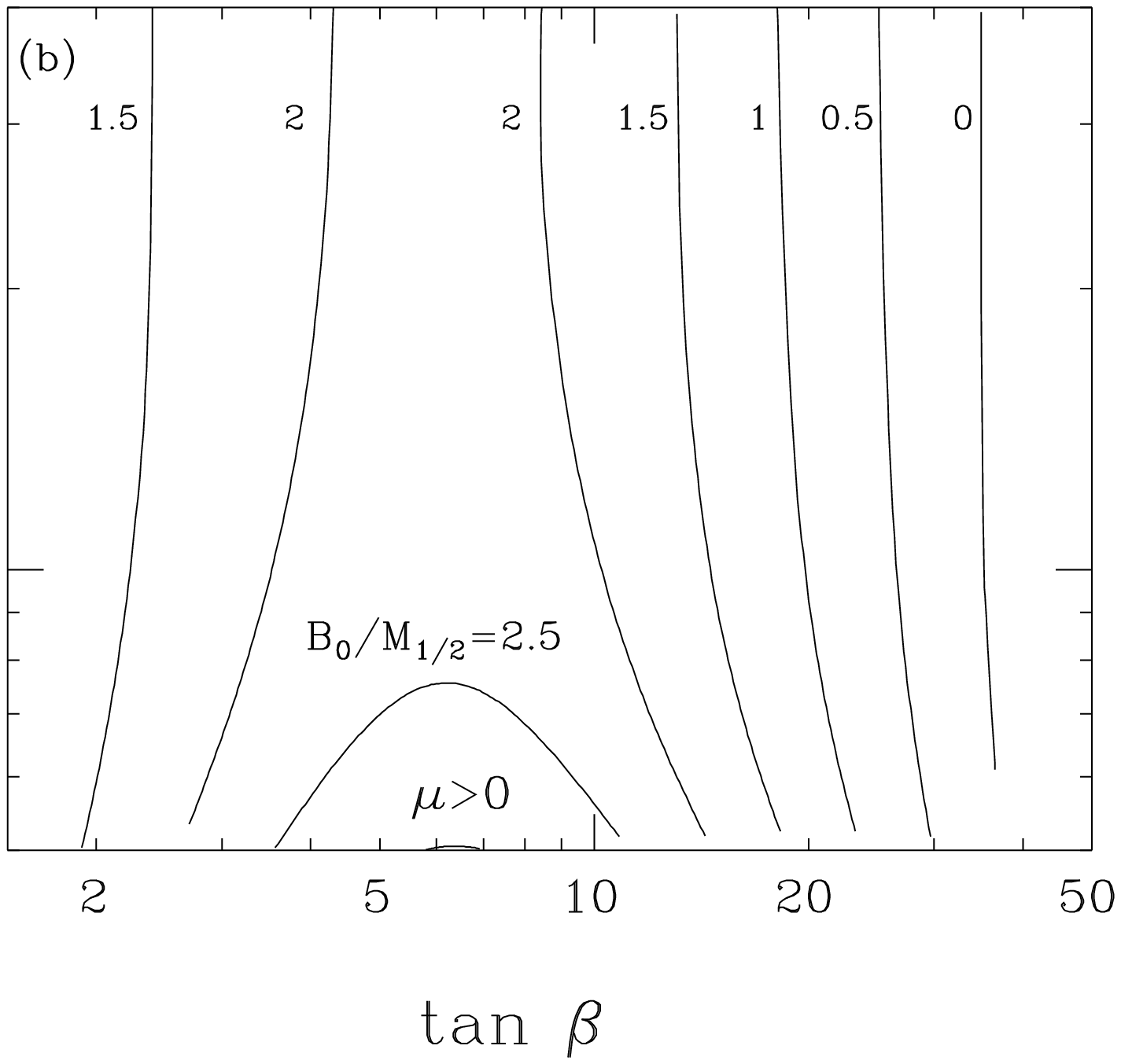}
\includegraphics{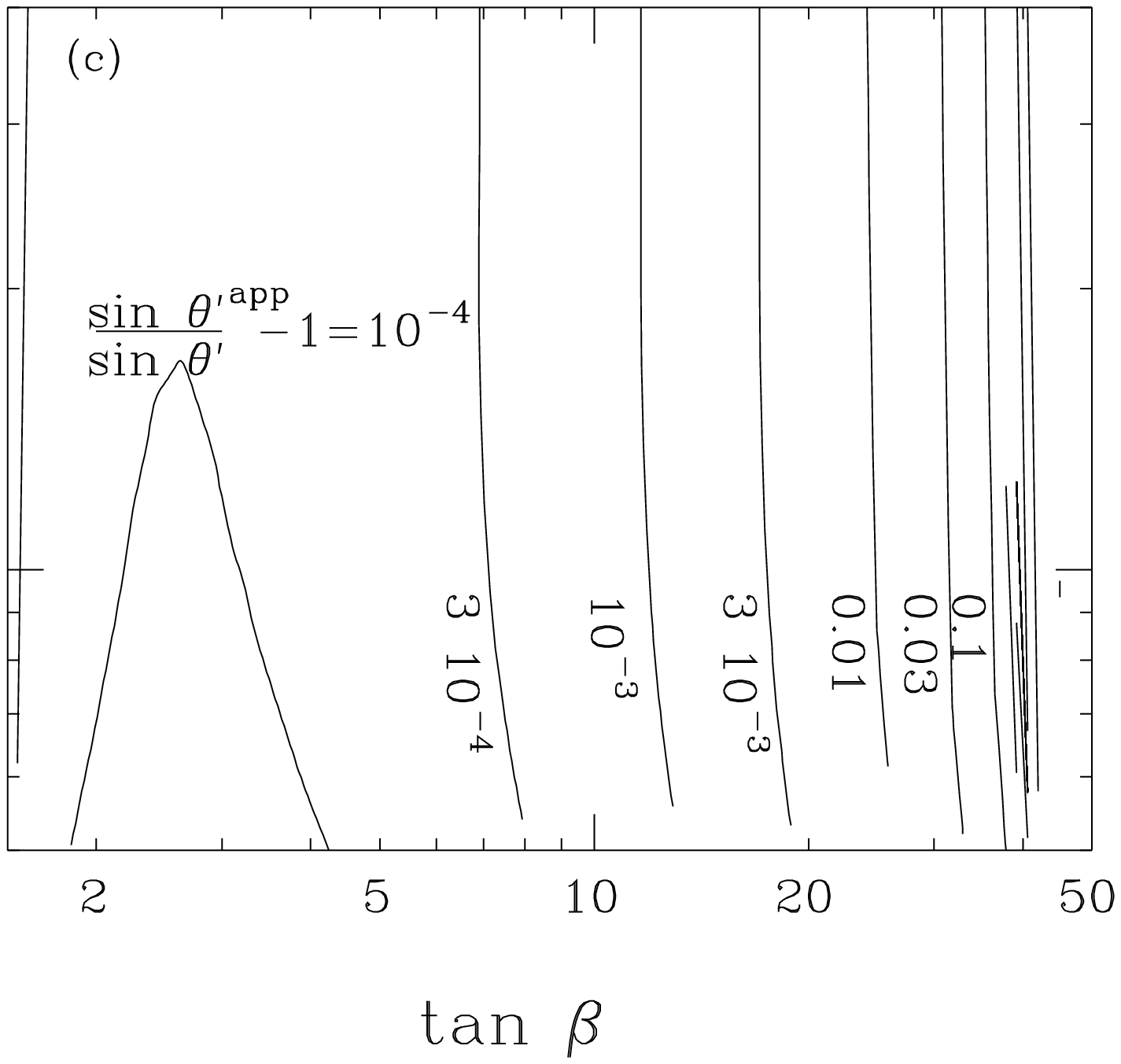}
\fcaption{Contours of constant $m_0$, $B_0$ and 
$\sin \theta_1^{\prime {\rm app}} / \sin \theta_1^{\prime}$
in the $\tan\beta$--$m_{1/2}$ plane.
We set $A_0 = 0$ and $\mu = 2.5 m_{1/2}$.
}
\label{fig1}
\end{figure}


The outline of our paper is as follows: in section~2 we present the
neutrino and sparticle spectrum obtained from REWSSB
including $\mu_i$. In section~3 
we present the numerical analysis of the 
LSP life-time, $\tau_{LSP}$, and LSP branching fractions.
Our conclusions are presented in section~4.

\section{Radiative Electro-Weak Symmetry Breaking}

Without any assumptions based on theoretical prejudice
there are many models with different 
SUSY particle spectra and vastly different
phenomenology.
Thus, it has become standard
to derive the low energy particle spectrum from
minimal supergravity model with only four independent parameters:
the universal scalar mass parameter, $m_0$,
the universal gaugino mass parameter, $m_{1/2}$,
and the universal $A$ ($B$) parameter multiplying the
tri-linear (bi-linear) terms in the superpotential [eq.~\ref{defw}].
This approach is supported by the observation
that the absence of FCNC implies a high mass-degeneracy of 
all scalars with the same gauge quantum numbers (with
the possible exceptions of
the Higgs mass parameters).

\begin{figure}
\vspace*{13pt}
\vspace*{3.6truein}      
\includegraphics{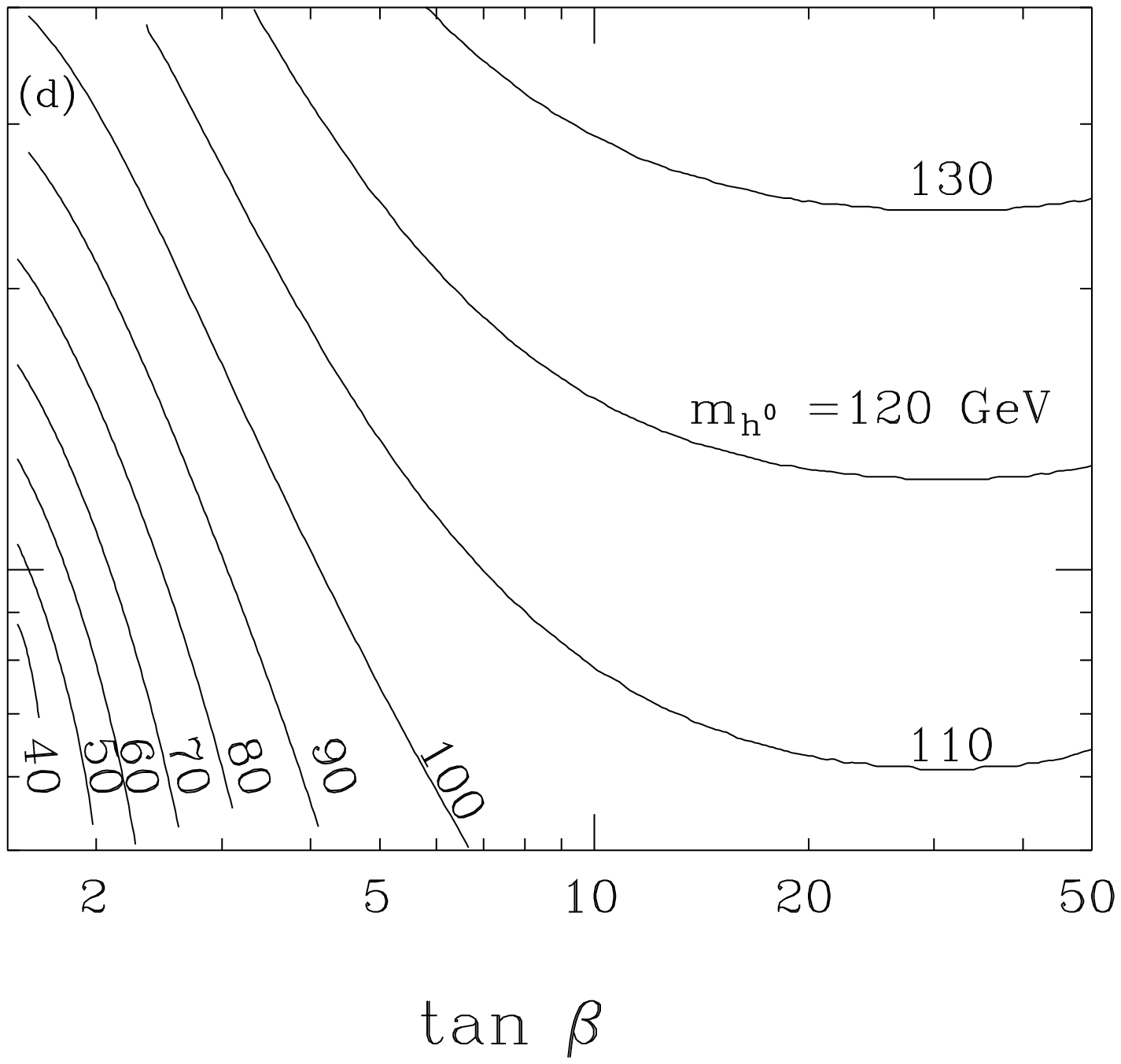}
\includegraphics{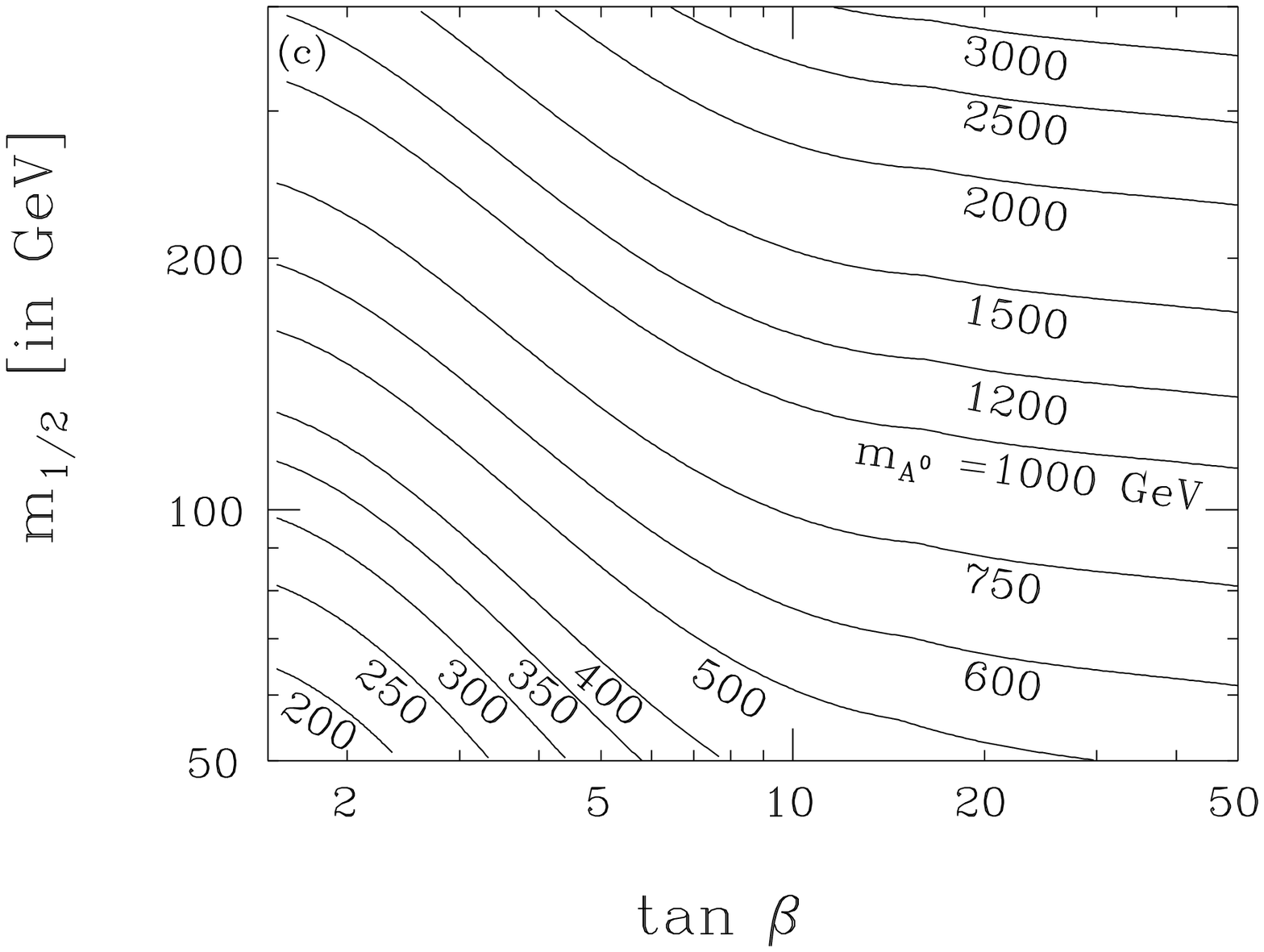}
\includegraphics{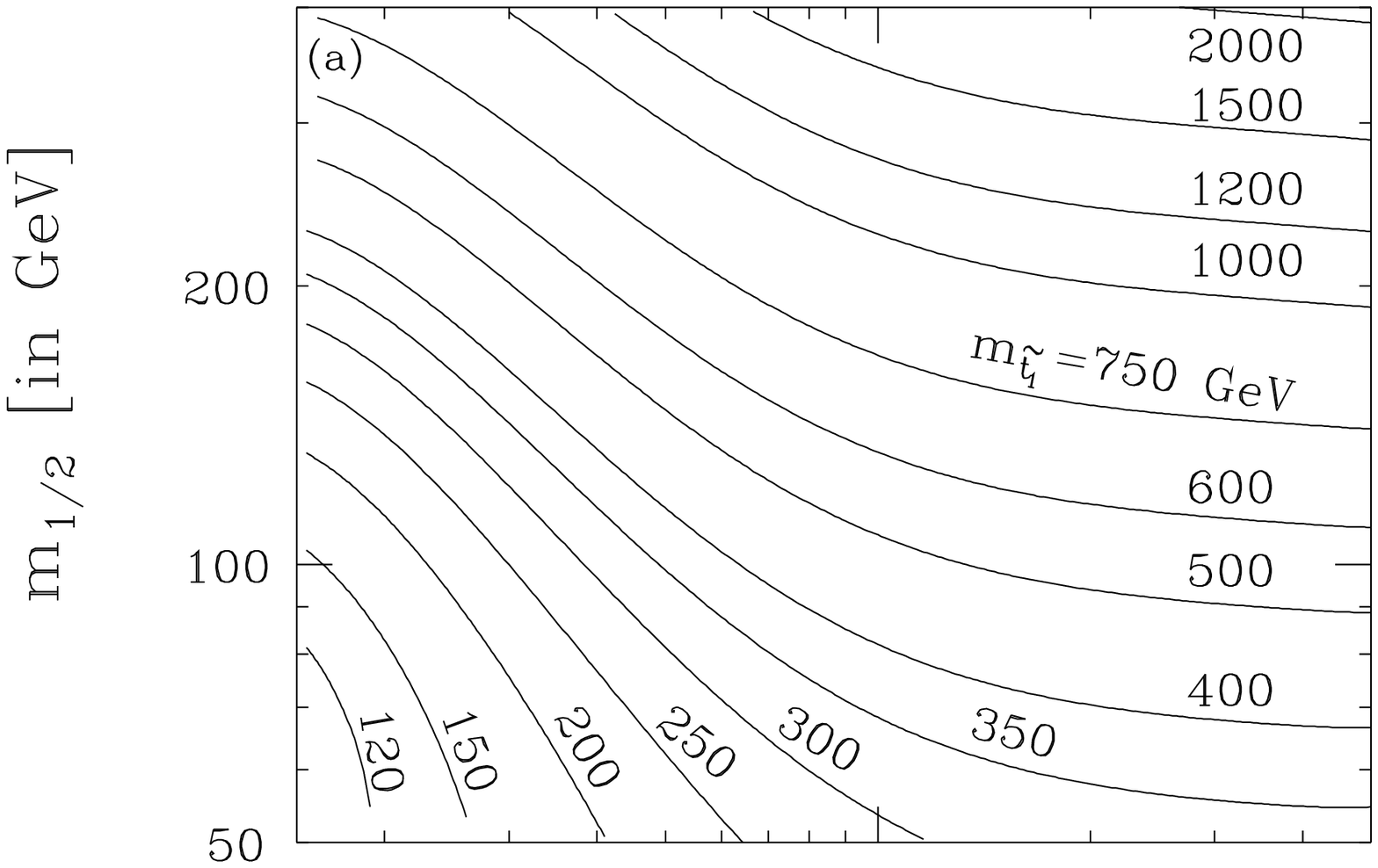}
\includegraphics{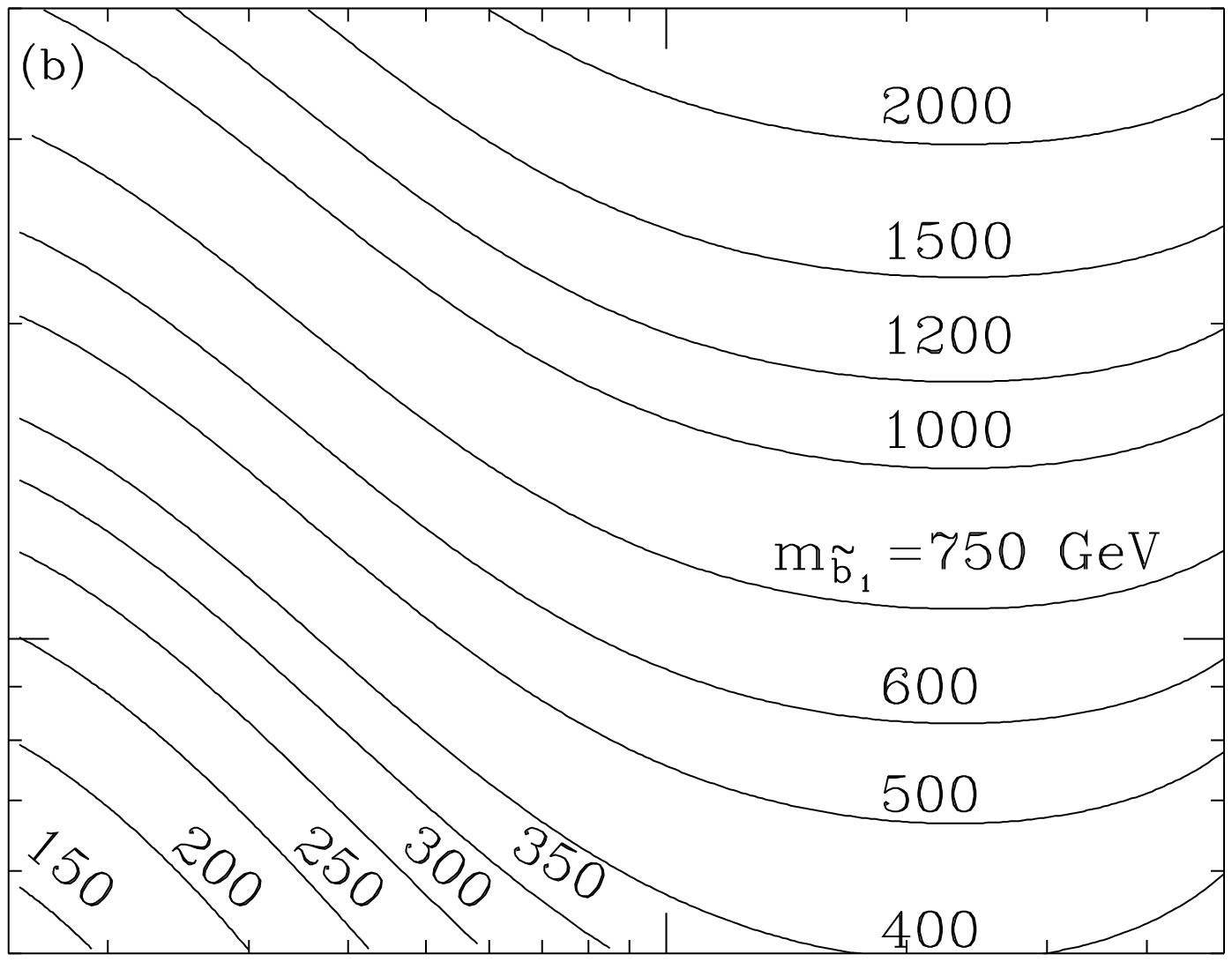}
\fcaption{Contours of constant
(a) $m_{\tilde t_1}$,
(b) $m_{\tilde b_1}$,
(c) $m_{A^0}$ (tree-level) and
(d) $m_{h^0}$ (incl. 1-loop radiative corrections)
in the $\tan\beta$--$m_{1/2}$ plane.
The other SUGRA parameters are as in fig.~\ref{fig1}.}
\label{fig2}
\end{figure}

First, we have to minimize the Higgs potential given by
\beqn
V &=& (\mu^2 + m_H^2) H^\dagger H
   +(\mu_I \mu_J + m_{L_{I J}}^2) \widetilde L_I^\dagger \widetilde L_J
   -B_{I J} \mu_I \left(\widetilde L_J H+\hc\right)\nonumber\\
   && +{g^2+g^{\prime 2}\over 8}\left(H^\dagger H 
          - \widetilde L^\dagger_I \widetilde L_I\right)^2
   +{g^2\over 2}\left|H^\dagger \widetilde L_I\right|^2\,,
\label{pot}
\eeqn
where the low energy soft SUSY breaking parameters are obtained
by renormalization group evolution below $\mgut$
in the standard fashion.
In order to stay as close to the notation of the MSSM as possible we
follow our notation of ref.~\citenum{npb478}
\beqn
\barv \equiv {\vev{H^0}\over \sqrt2}\,,~
v_I   \equiv {\vev{\widetilde L^0_I} \over \sqrt2}\,,~
v     \equiv  \sqrt{v_I v_I}\,,~\hbox{and}  ~
\tanb \equiv \barv/v\,,
\eeqn
and we parameterize the VEVs in terms of spherical
coordinates
\beqn
\tan\theta_3^\prime = {v_3\over v_2}\,,\qquad
\tan\theta_2^\prime = {v_2\over v_1 \cos \theta_3}\,,\qquad
\tan\theta_1^\prime = {v_1\over v_0 \cos \theta_2}\,.
\label{deftani}
\eeqn
Analogously, it is convenient to parameterize the $R_p$ breaking
mass parameters in terms of three mixing angles
\beqn
\tan\theta_3 = {\mu_3\over \mu_2}\,,\qquad
\tan\theta_2 = {\mu_2\over \mu_1 \cos \theta_3}\,,\qquad
\tan\theta_1 = {\mu_1\over \mu_0 \cos \theta_2}\,,
\label{deftaniprime}
\eeqn
and $\mu \equiv \sqrt{\mu_I \mu_I}$.
The potential in eq.~\ref{pot} is minimized by an iterative procedure
using the analytic solution for $\tan \theta_1 = 0$ 
as our initial values.
This procedure also works surprisingly well for $\tan \theta_1 >1$.
For small $R_p$ violating
parameters we can also obtain very reliable analytic expressions
in the basis where $y^L_{i j}$ is diagonal
by expanding in powers of $\mu_i/\mu_0$
\beqn
\sin 2 \beta &=& 
         {2 B_{00} \mu_0 \over m_{L_{00}}^2+m_H^2+2 \mu_0^2}\,,\label{sin2b} \\
\tan^2 \beta &=& {m_{L_{00}}^2+ \mu_0^2+\half \mz^2
                 \over m_H^2  + \mu_0^2+\half \mz^2}\,,\label{tanb2}\\
{v_i\over v_0}&=& \mu_i {B_{(i i)} \tanb - \mu_0
                    \over m_{L_{(i i)}}^2+\half \mz^2 \cos 2\beta}\,,
\label{viv0}
\eeqn
with the convention that indices in braces are not summed over.
In general, one fixes the GUT input parameters are $m_0$, $m_{1/2}$ and $A_0$.
$B_{0 0}$ and $\mu$ are obtained by solving eq.~\ref{sin2b} and \ref{tanb2}
while keeping $\tanb$ and $v$ fixed.
Here, we find it convenient to fix the fermionic spectrum
given by $\mu$ rather than the scalar spectrum determined by $m_0$.

\begin{figure}
\vspace*{13pt}
\vspace*{2.1truein}      
\includegraphics{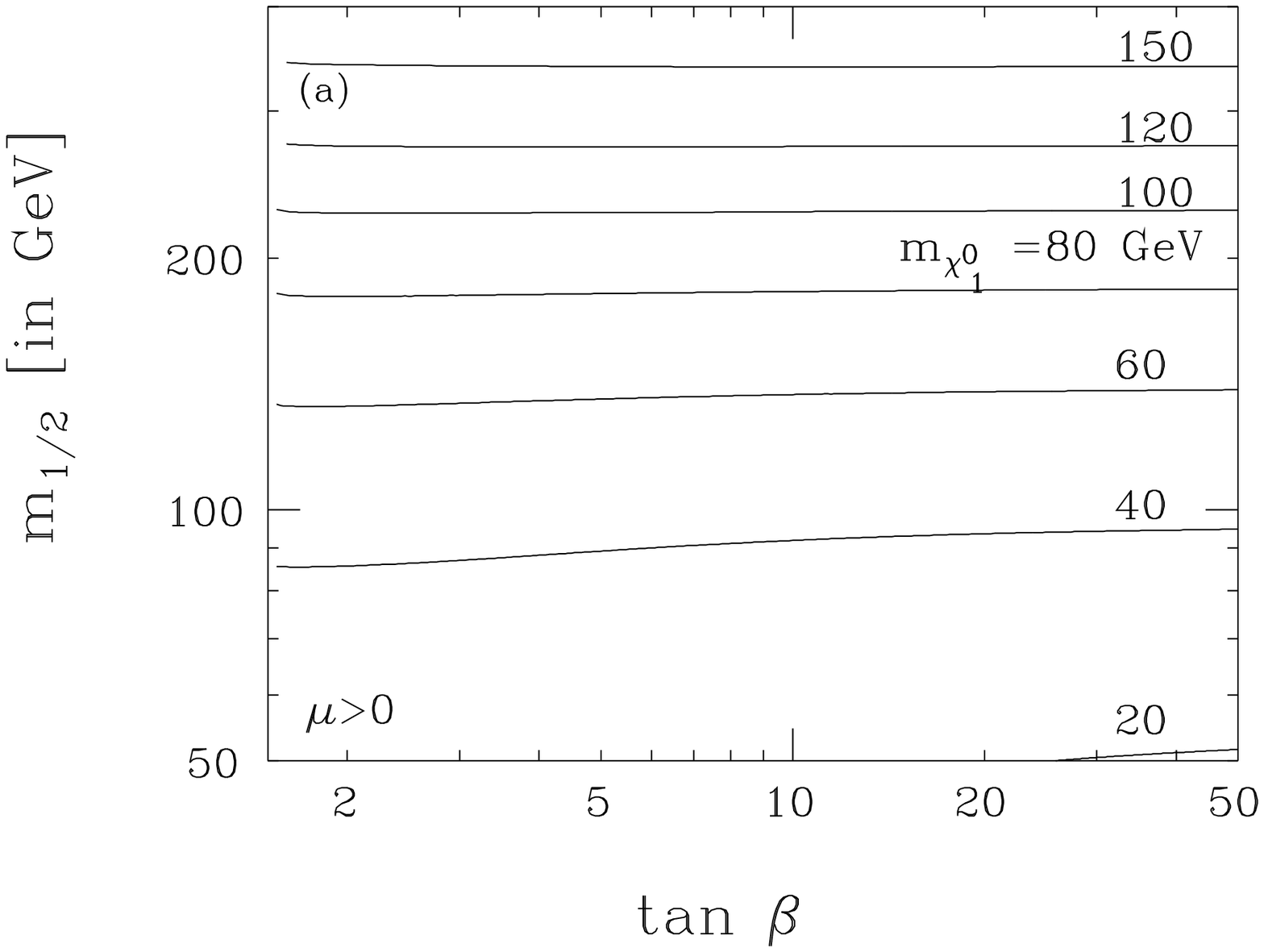}
\includegraphics{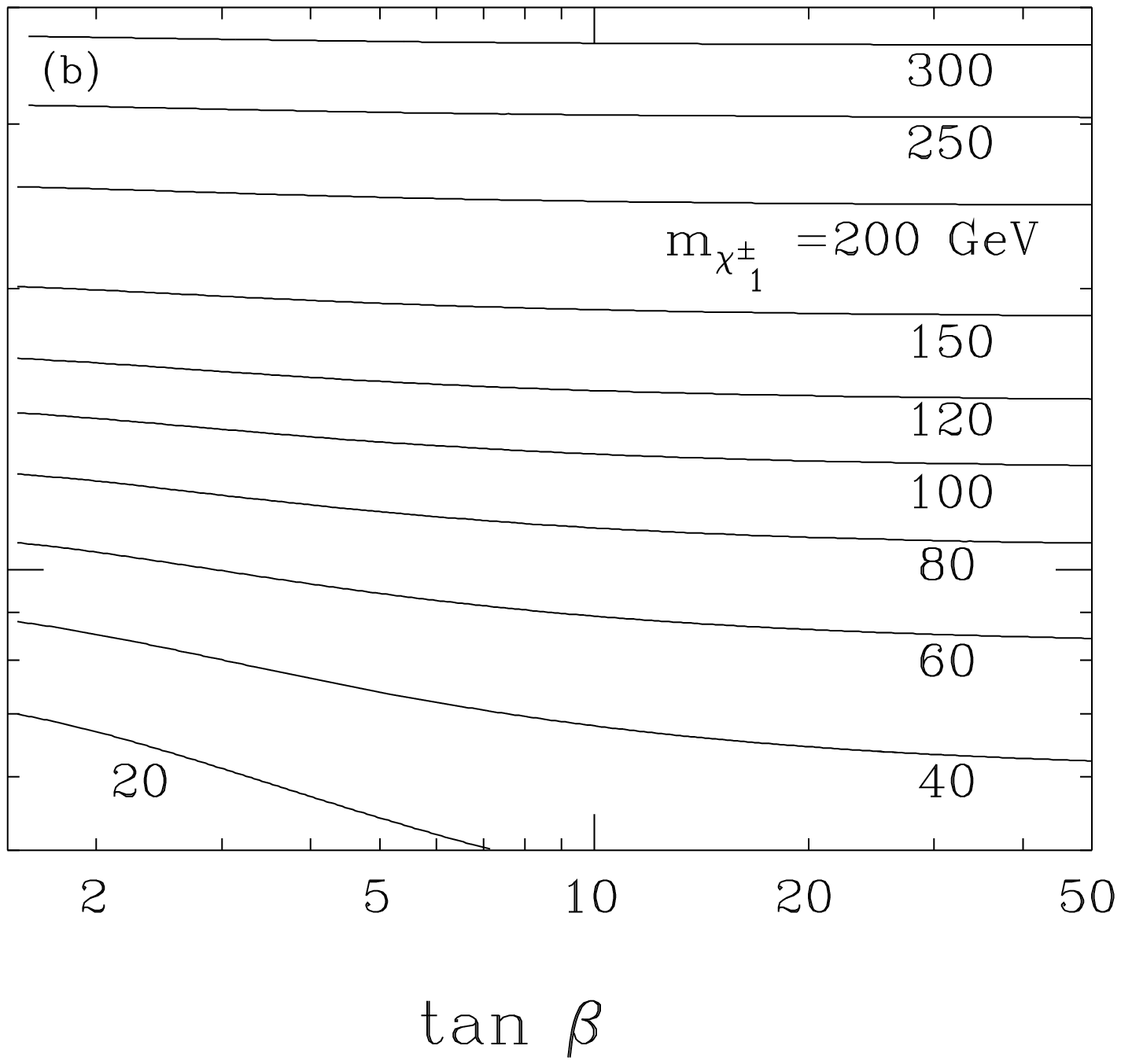}
\fcaption{Contours of constant
(a) $m_{\chi_1^0}$,
(b) $m_{\chi_1^\pm}$,
in the $\tan\beta$--$m_{1/2}$ plane.
The other SUGRA parameters are as in fig.~\ref{fig1}.
}
\label{fig3}
\end{figure}

In fig.~\ref{fig1}(a) and (b) we present contours of
constant GUT parameters $m_0$ and $B_0$
in the $\tanb$--$m_{1/2}$ plane.
and in fig.~\ref{fig1}(c) we show how well the approximation works
for the minimization of the potential (eq.~\ref{viv0}).
We see that the deviation of the approximation
obtained from eq.~\ref{viv0} 
denoted by $\sin \theta_1^{\prime {\rm app}}$
from the true minimum
obtained by numerical methods and denoted by $\sin \theta_1^\prime$
is quite small as long as
$\tanb = O(1\sim 10)$. However, it breaks down
for $\tanb \gsim 40$.

\subsection{Sparticle Spectrum}

From LEP experiments we know that there are no
charged superpartners with mass below $\mz/2$.
Furthermore, we can deduce a similar constraint
on the lightest neutralino which, in our model, is instable.
In fig.~\ref{fig2} and~\ref{fig3} we present
contours of some relevant scalar and fermionic
superpartner masses in the $\tanb$--$m_{1/2}$ plane.
We have chosen $A_0=0$ and $\mu = 2.5 m_{1/2}$.
The value of $m_0$ is obtained by imposing REWSSB
[see fig.~\ref{fig1}(a)]. We see that 
the only relevant constraint arises from the experimental lower
on the lightest neutralino mass
denoted by $M_{LSP}$.

\begin{figure}
\vspace*{13pt}
\vspace*{3.6truein}      
\includegraphics{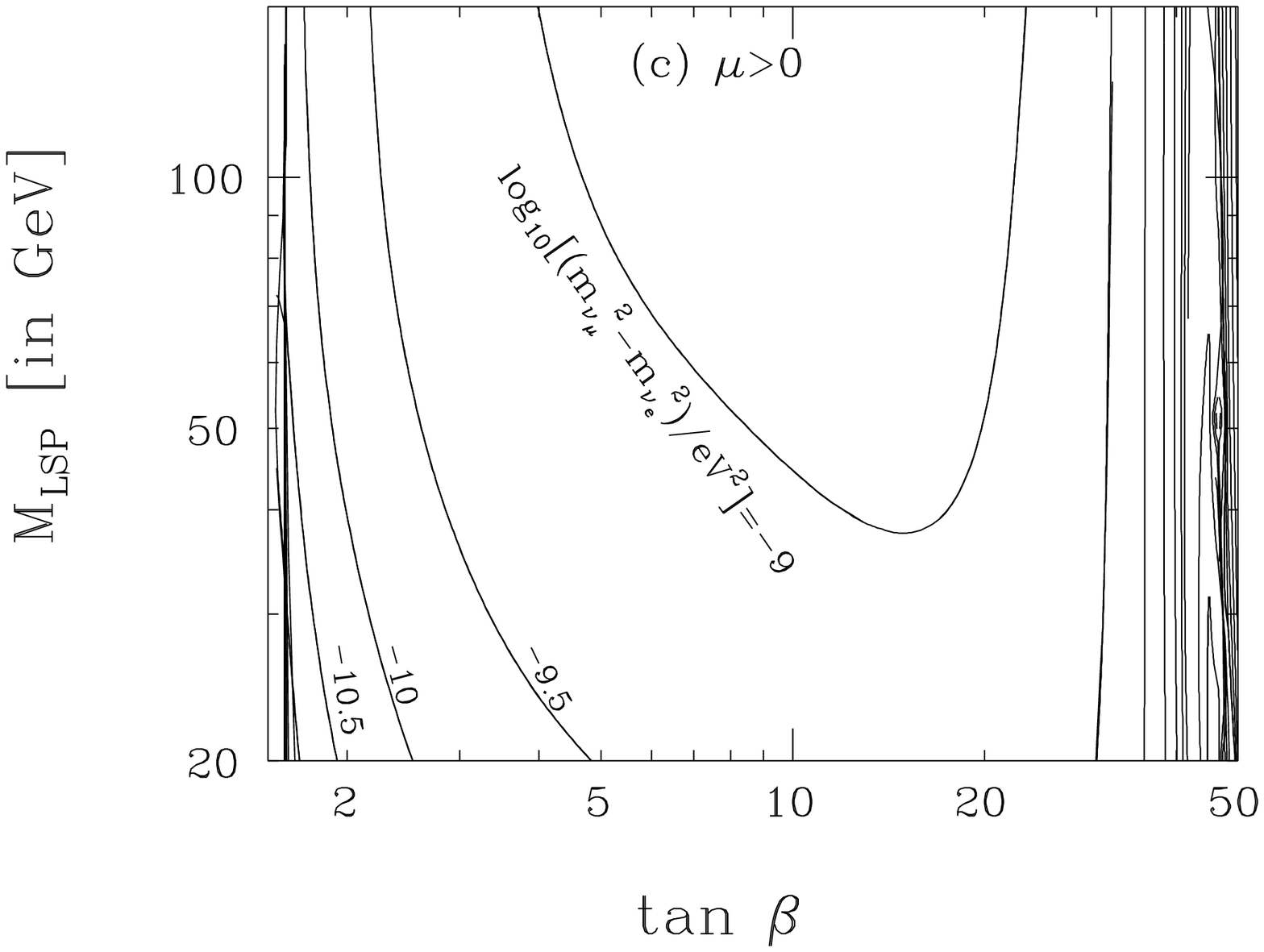}
\includegraphics{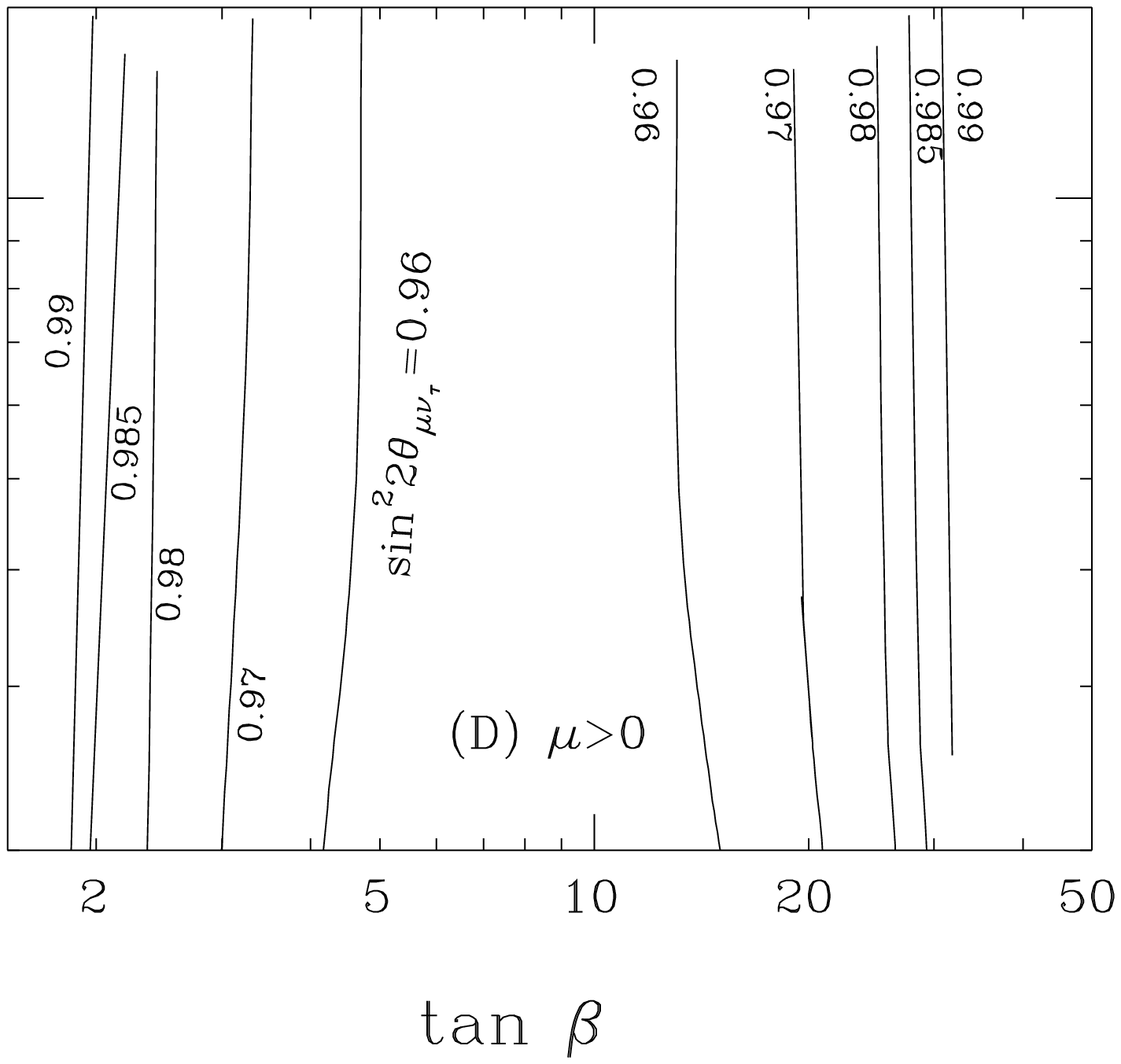}
\includegraphics{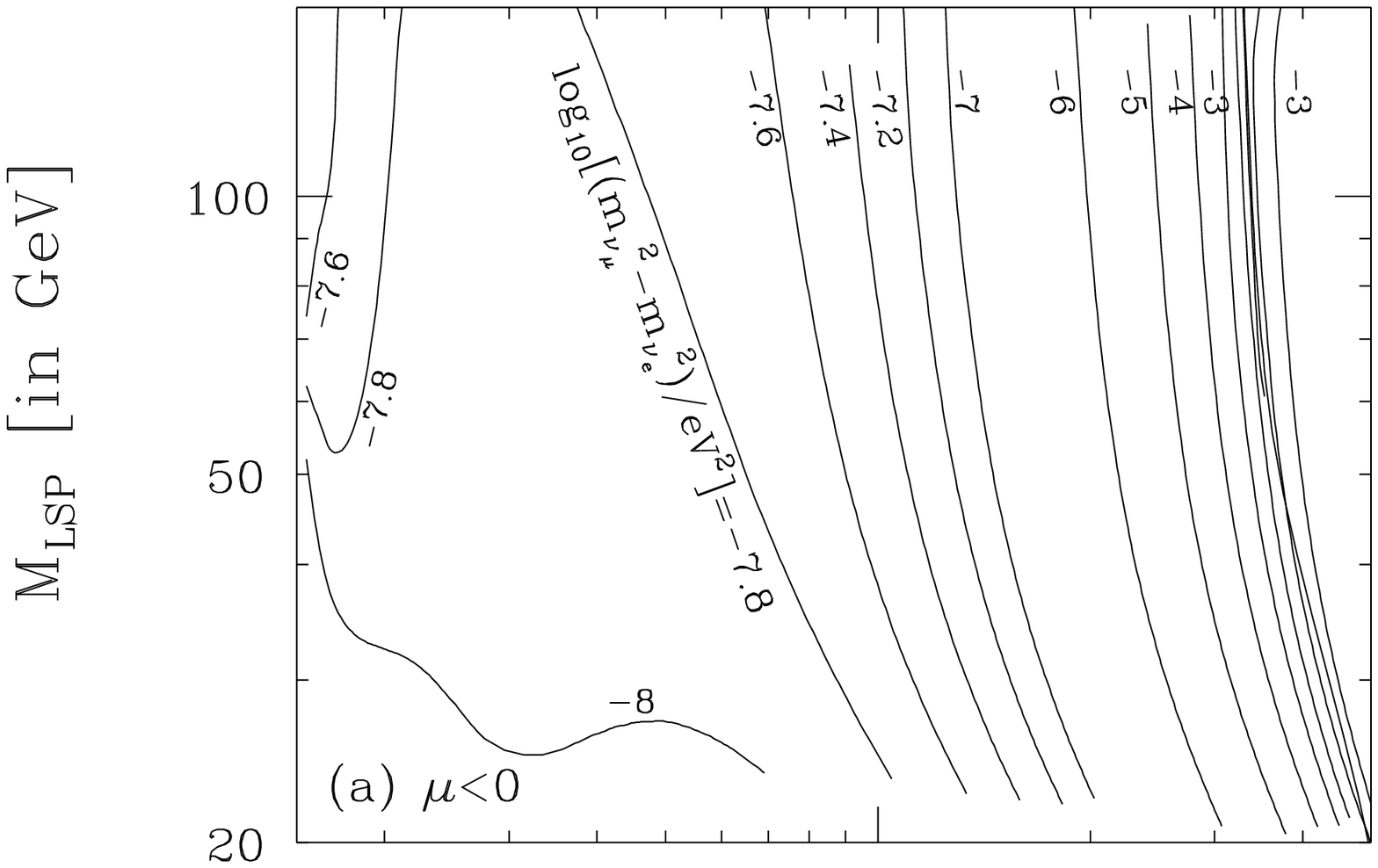}
\includegraphics{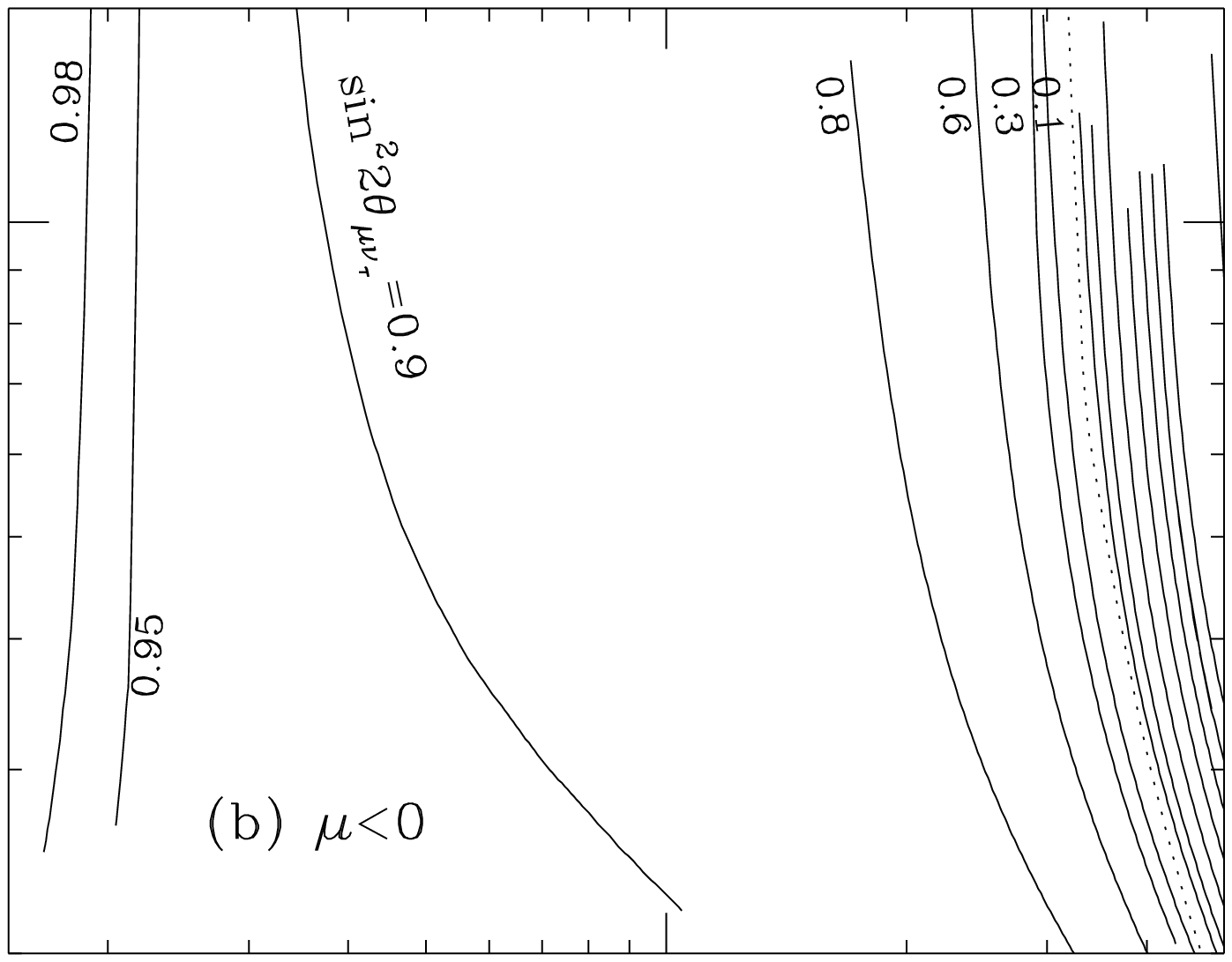}
\fcaption{Contours of constant
(a) $m_{\nu_\mu}^2-m_{\nu_e}^2$ 
(b) $\sin^2 2\theta_{\tau \nu_\mu}$
in the $\tan\beta$--$M_{LSP}$ plane.
We set $A_0 = 0$ and $\mu = 2.5 m_{1/2}$ and $m_{1/2}$ is replaced
in favor of $M_{LSP}$.
}
\label{fig4}
\end{figure}

\subsection{Neutrino Spectrum}

The LSP phenomenology of the model under investigation here is 
governed to a very good approximation by only one parameter,
$\tan \theta_1$. This parameter also determines the 
neutrino masses whose upper limits are given by
\beqn
m_{\nu_e}  \leq 4.35~{\rm eV}\,,
m_{\nu_\mu}  \leq  160~{\rm keV}\,,
m_{\nu_\tau} \leq  23~{\rm MeV}\,,
\qquad
&&\hbox{Collider-experiment~\cite{nmasses}}
\nonumber\\
\sum_{x=e,\mu,\tau} m_{\nu_x} \leq (10\sim 100)~{\rm eV} \,,
\qquad\qquad\qquad\qquad\qquad\qquad
&&\hbox{Cosmology~\cite{k&t}}
\label{nutau-limits}
\eeqn
While the is no direct evidence for non-zero neutrino masses,
there is strong experimental evidence for 
neutrino oscillations which imply that the three neutrinos are non
mass-degenerate.
In this work, we will take the view that our $R_p$ violating 
terms are the only source of neutrino masses
and, therefore, should account for all the existing neutrino mixing
effects.
Since not all experimental 
indication for neutrino oscillations appear compatible
with each other or have the same statistical significance
a selection has to be taken.\footnote{%
It has recently been suggested that
three neutrino flavor are enough to
accomodate all three indication for neutrino ocillation\cite{acker}.
However, for the sake of generality we will
be more conservative.}
The solar neutrino puzzle\cite{solarn} may be the most
compelling evidence for neutrino mixing.
However, the effect appears to involve only
the first two neutrino flavors neither of which
is likely to be the heaviest neutrino both on theoretical and experimental
grounds. Thus, the solar neutrino puzzle is not 
well suited for a determination of $\tan \theta_1$.
Instead we choose to solve the atmospheric neutrino
problem\cite{atmosphericn}.
This requires that we fix 
$\tan \theta_1$ such that
$m_{\nu_\tau}^2 - m_{\nu_\mu}^2 = 10^{-2}$~eV$^2$ and we set
$\tan \theta_2 = 1$ (we use a small value for $\tan \theta_3 = 0.045$
in order to solve the solar neutrino puzzle via the
MSW-effect\cite{msw-effect};
this angle will turn out to be quite irrelevant otherwise).
Over most of the parameter space under consideration here
this implies $ m_{\nu_\mu}\ll m_{\nu_\tau} \simeq 0.1$~eV.
It is then straightforward to obtain lower limits on $\tau_{LSP}$
from upper limits on $m_{\nu_\tau}$ [eq.~\ref{nutau-limits}]
by simple scaling arguments.

\begin{figure}
\vspace*{13pt}
\vspace*{2.1truein}      
\includegraphics{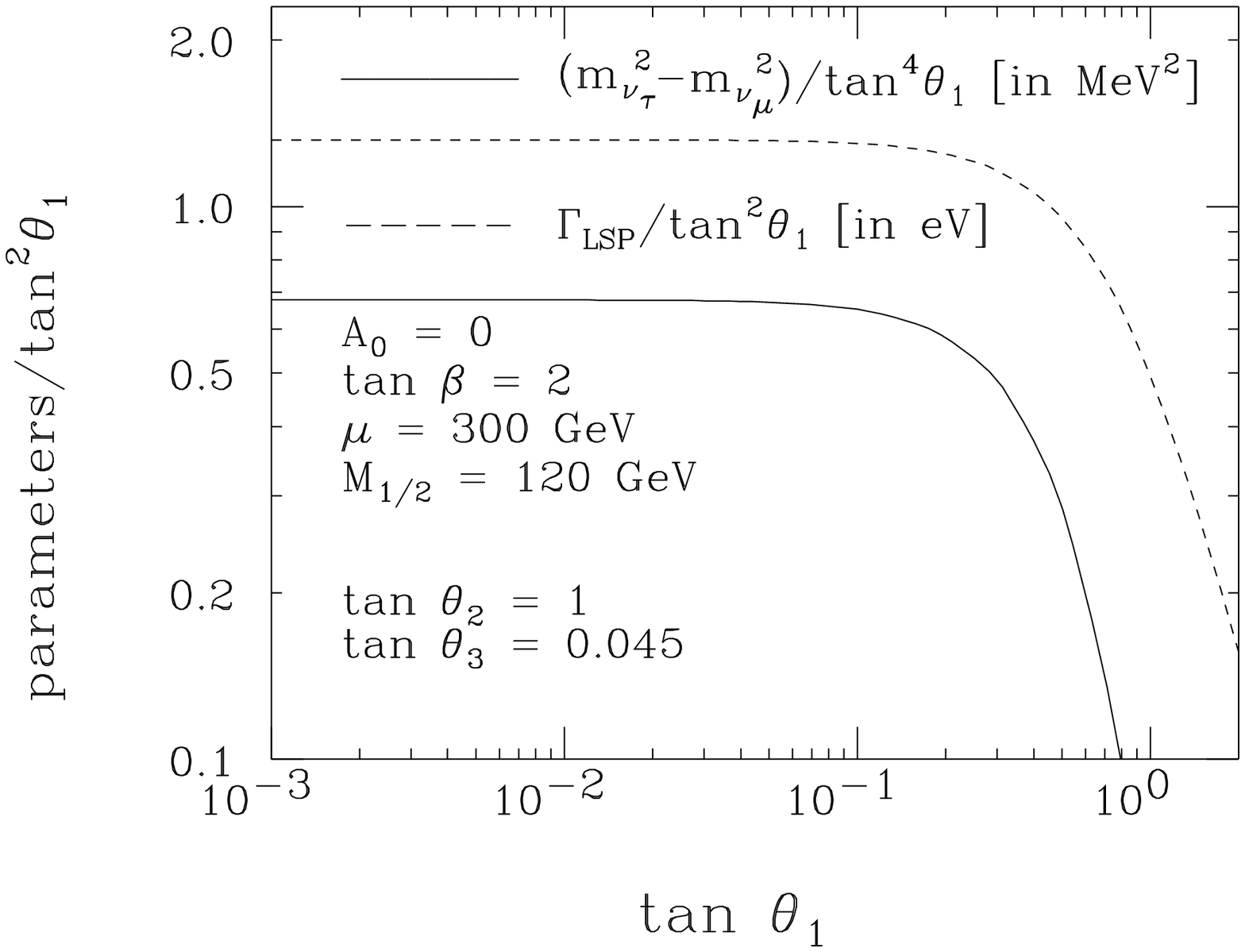}
\includegraphics{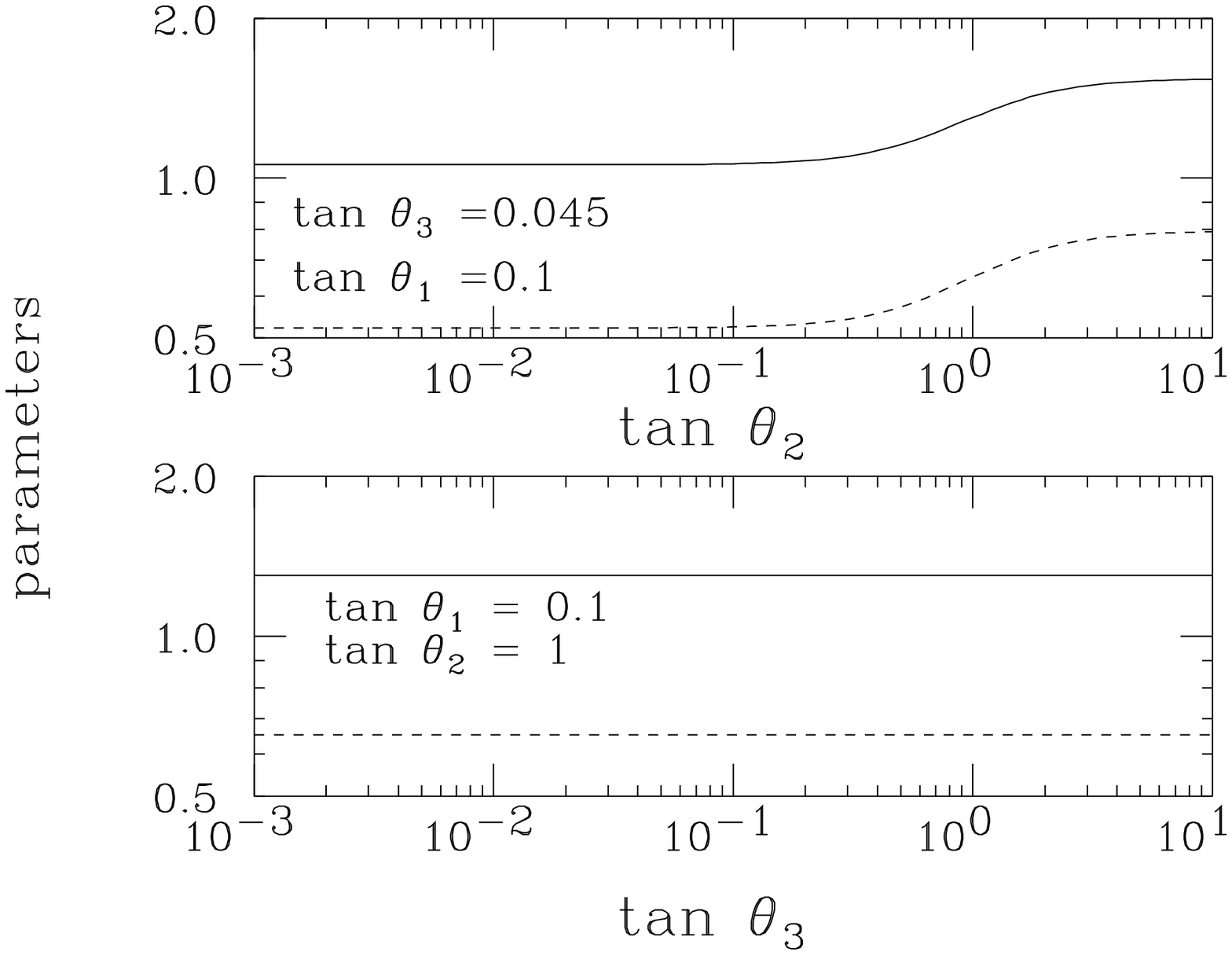}
\fcaption{The LSP-width, $\Gamma_{LSP}$
(divided by tan$^2\theta_1$)
and the mass difference $m_{\nu_\tau}^2 - m_{\nu_\mu}^2$
(divided by tan$^4\theta_1$)
as a function of
tan$\theta_1$, tan$\theta_2$ and tan$\theta_3$.
}
\label{fig5}
\end{figure}

In fig.~\ref{fig4} we present contours of constant
values for $m_{\nu_\mu}^2 - m_{\nu_e}^2$
and sin$^2 2\theta_{\mu \nu_\tau}$. We fix $\mu = \pm 2.5 m_{1/2}$
and $A_0=0$.
We see that for positive values of $\mu$ the mass 
difference $m_{\nu_\mu}^2 - m_{\nu_e}^2 = O(10^{-9}~\ev^2)$
is very small and quite compatible with long wave oscillation (LWO)\cite{lwo}
solution to the solar neutrino problem [fig.~\ref{fig4}
does not change if we set tan$\theta_3 = O(1)$].
For $\mu<0$ there is a sizable region were
$m_{\nu_\mu}^2 - m_{\nu_e}^2 = O(10^{-5}~\ev^2)$
as required by the MSW explanation of the solar neutrino
deficiency\cite{solarn}.
These results were first presented in ref.~\citenum{npb478}\footnote{%
Note that fig. 7(a), fig. 8(a) and fig. 9 are mislabeled
in ref.~\citenum{npb478}. The region with $\mu>0$ and 
the region with $\mu<0$ should be interchanged.}

\begin{figure}
\vspace*{13pt}
\vspace*{3.6truein}      
\includegraphics{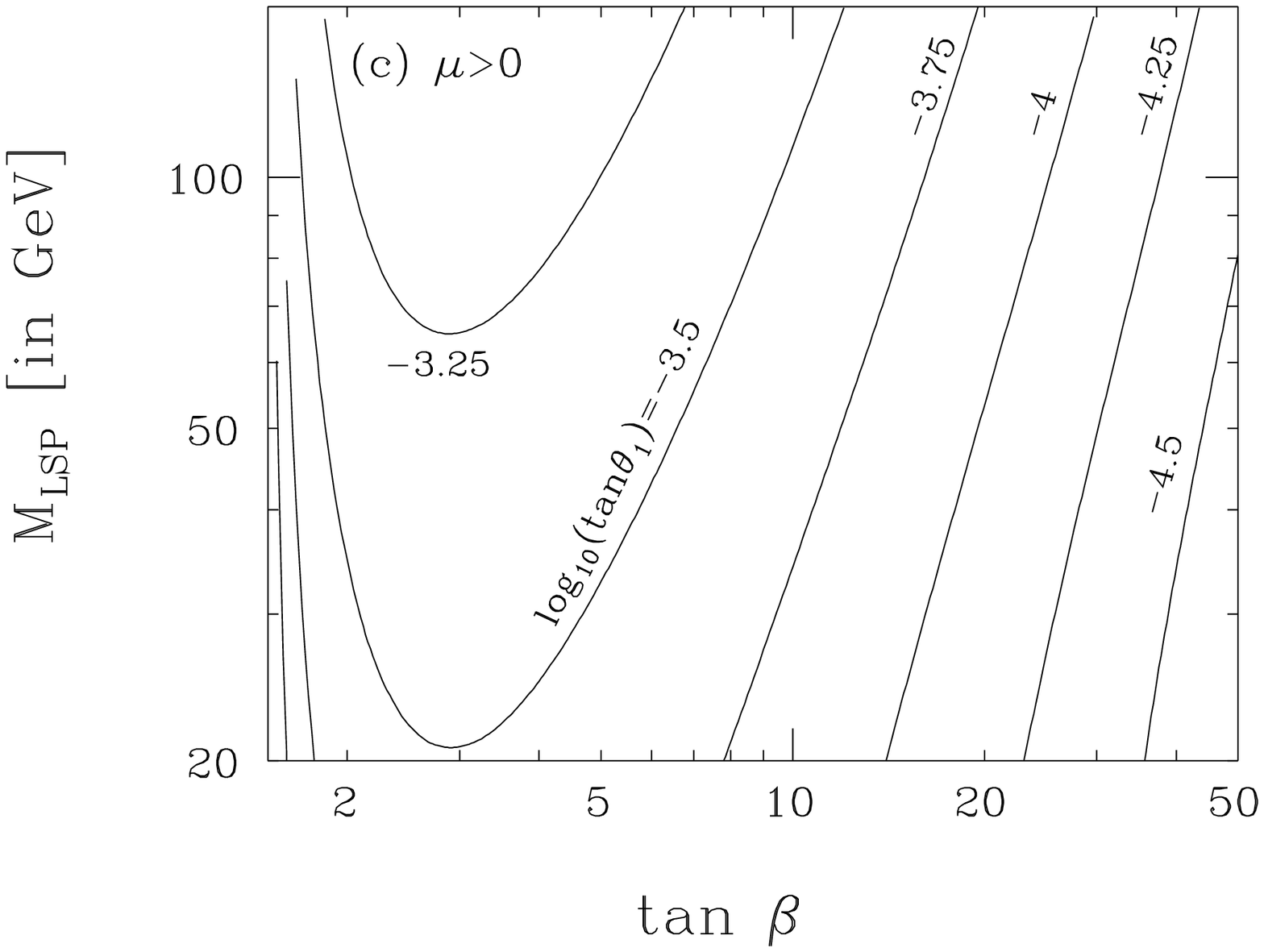}
\includegraphics{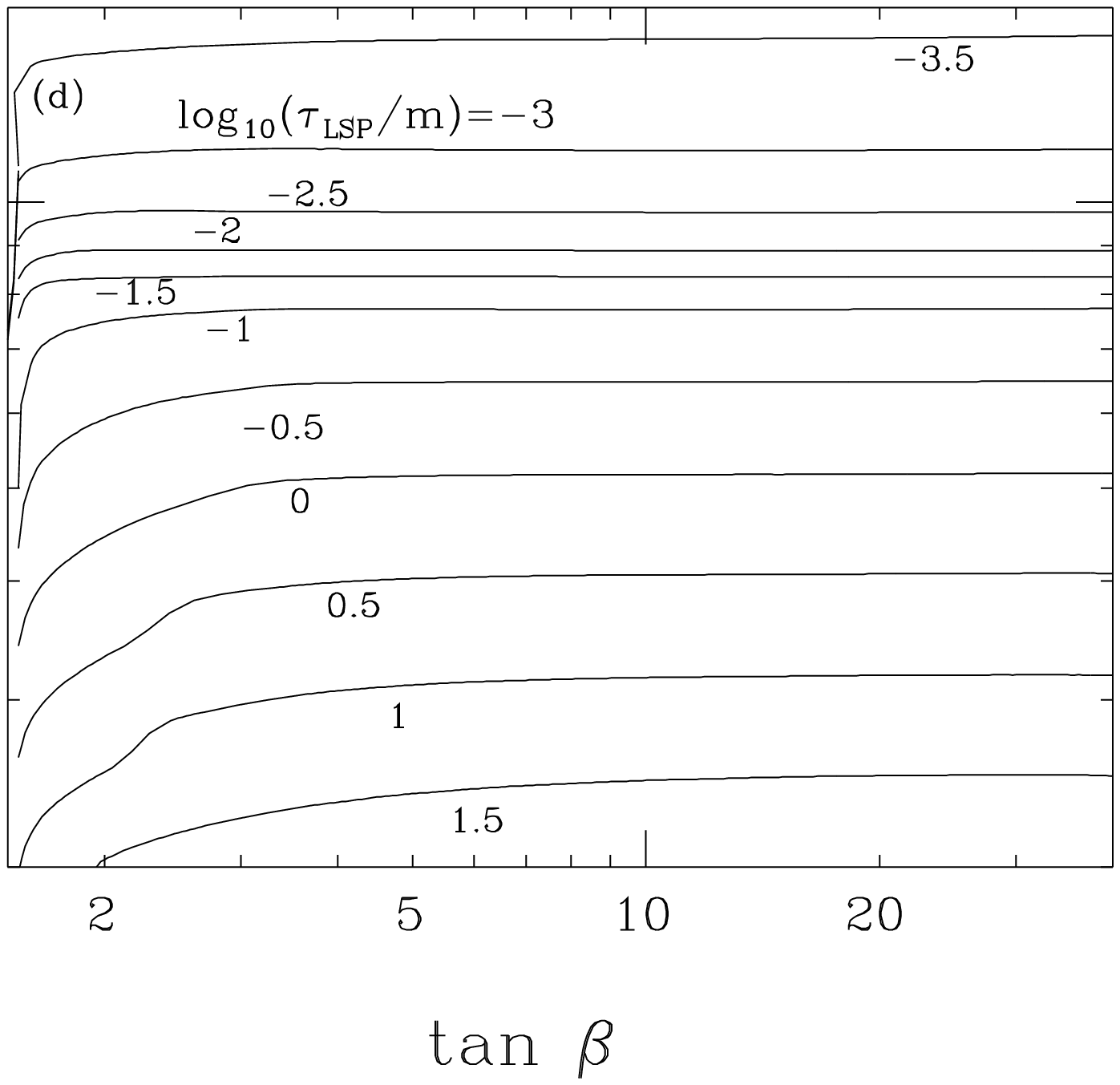}
\includegraphics{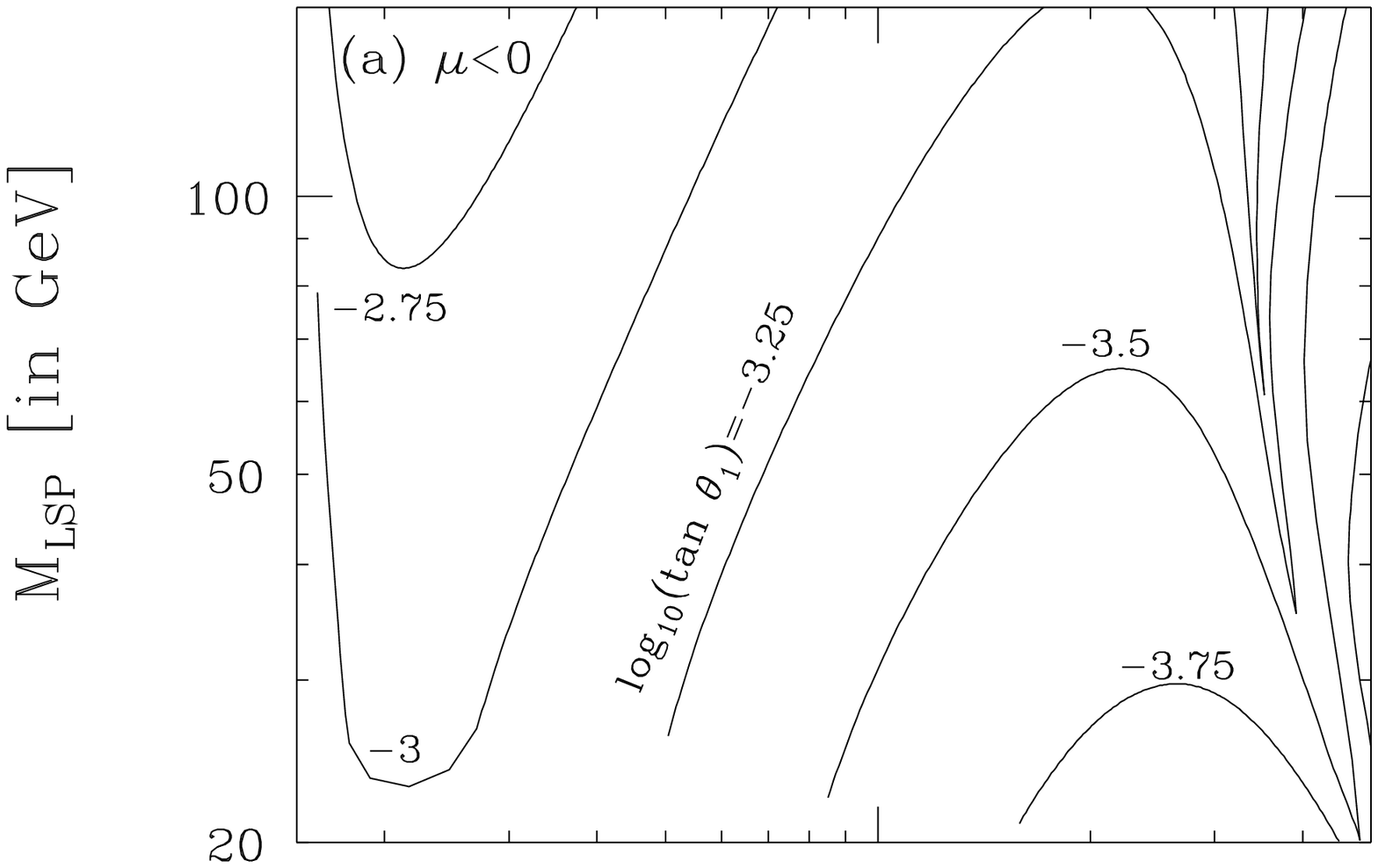}
\includegraphics{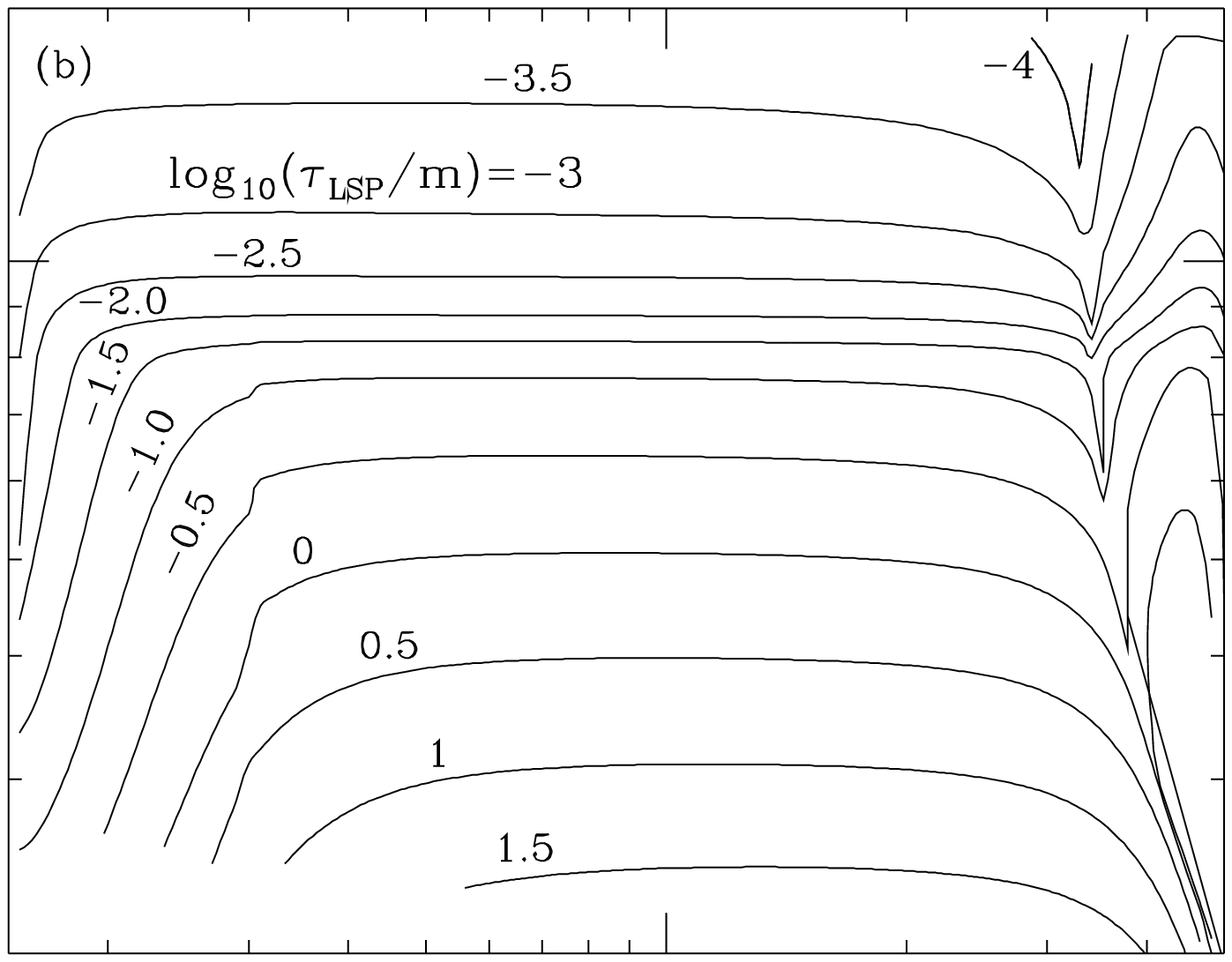}
\fcaption{Contours of constant
$\tan \theta_1$ and
$\tau(\chi^0_1)$
in the $\tan\beta$--$M_{LSP}$ plane for $\mu = \pm 2.5 m_{1/2}$.
The other SUGRA parameters are as in fig.~\ref{fig4}.}
\label{fig6}
\end{figure}

\section{LSP Phenomenology}

In this section, we will discuss the decay properties of the LSP.
The main interest from the point of view of collider phenomenology is
the question whether the LSP decays inside the detector
(else the analysis is equivalent to the case of unbroken
$R_p$).

Since the magnitude of $R_p$ violation is parameterized by
an priori free parameter $\tan\theta_1$, we cannot determine
$\tau_{LSP}$. The situation changes if we relate
$\tan\theta_1$ to the neutrino masses.
In the first part of this section we
will present the prediction for $\tau_{LSP}$ assuming
the atmospheric neutrino puzzle is a result of $R_p$ violation.
[This prediction can be turned into a lower limit
by imposing any of the  upper limits on $m_{\nu_\tau}$
of eq.~\ref{nutau-limits}.]
In the second part of this section, we will discuss the
branching fractions of the LSP which is
independent of $\tan\theta_1$ and, hence,
also of any assumption about neutrino masses.

\subsection{LSP life-time}

As pointed out in ref.~\citenum{suzuki}
in models without $R_p$ the neutrinos and neutralinos are
indistinguishable. As a result,
the formul\ae\ for the neutralino radiative decay
in the MSSM~\cite{wyler}
and the decay into three fermions~\cite{majerotto} can be 
directly generalized to our model.
However, we do have to include the effects of Yukawa couplings
which were neglected for $R_p$ preserving three-body decays~\cite{majerotto}.
Our complete set of formul\ae\ will be presented elsewhere\cite{formulae}.
Here, we simply present our numerical results.

\begin{figure}
\vspace*{3pt}
\vspace*{2.1truein}      
\includegraphics{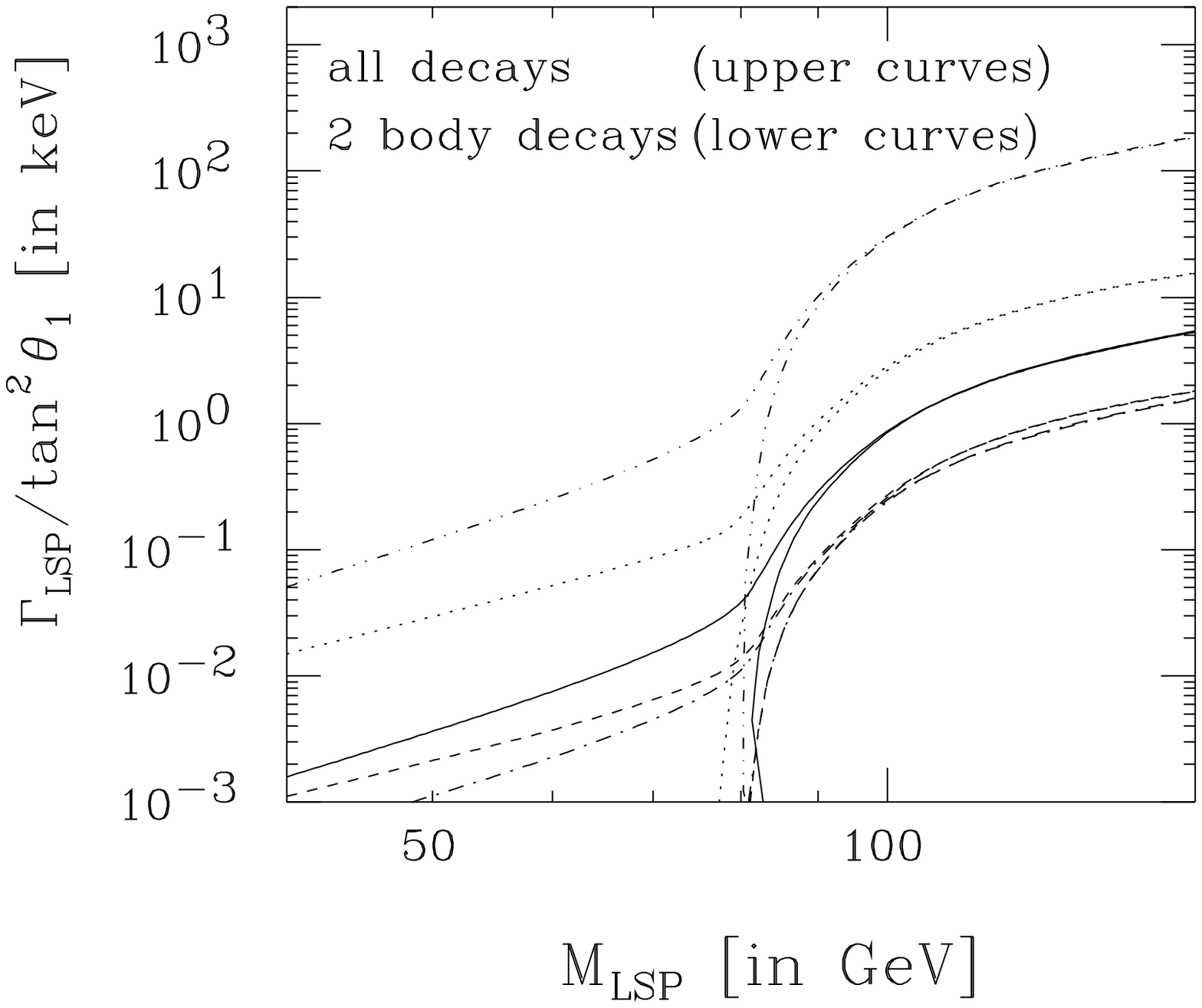}
\includegraphics{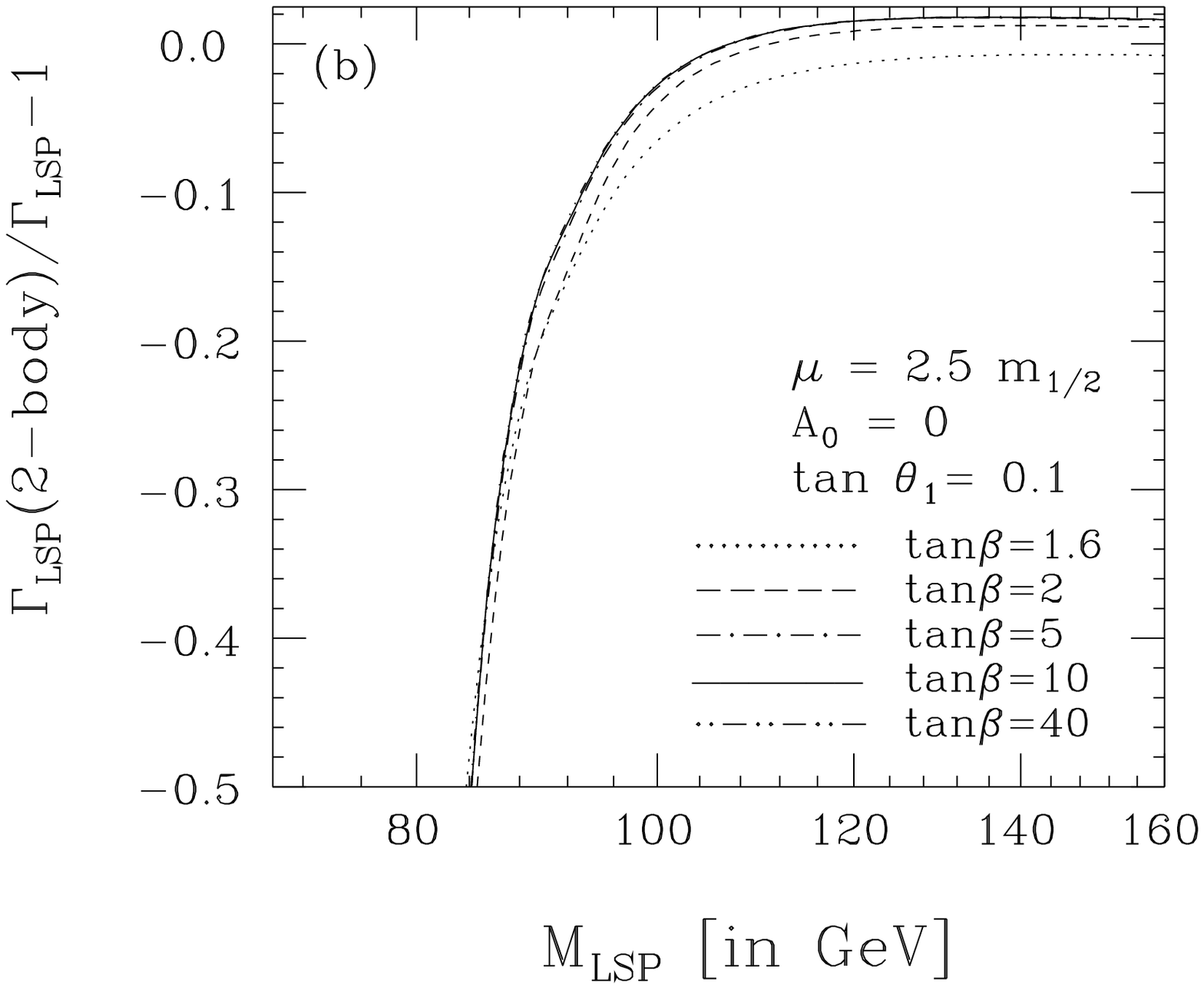}
\fcaption{
A comparison of the total LSP-width
with the  LSP-width due to two-body decays.
In (a) we show both sets of curves
as a function of $M_{LSP}$.
In (b) we show the difference of total width minus two-body decays
normalized to $\Gamma_{LSP}$.
}
\label{fig7}
\end{figure}

\begin{figure}
\vspace*{3pt}
\vspace*{2.1truein}      
\includegraphics{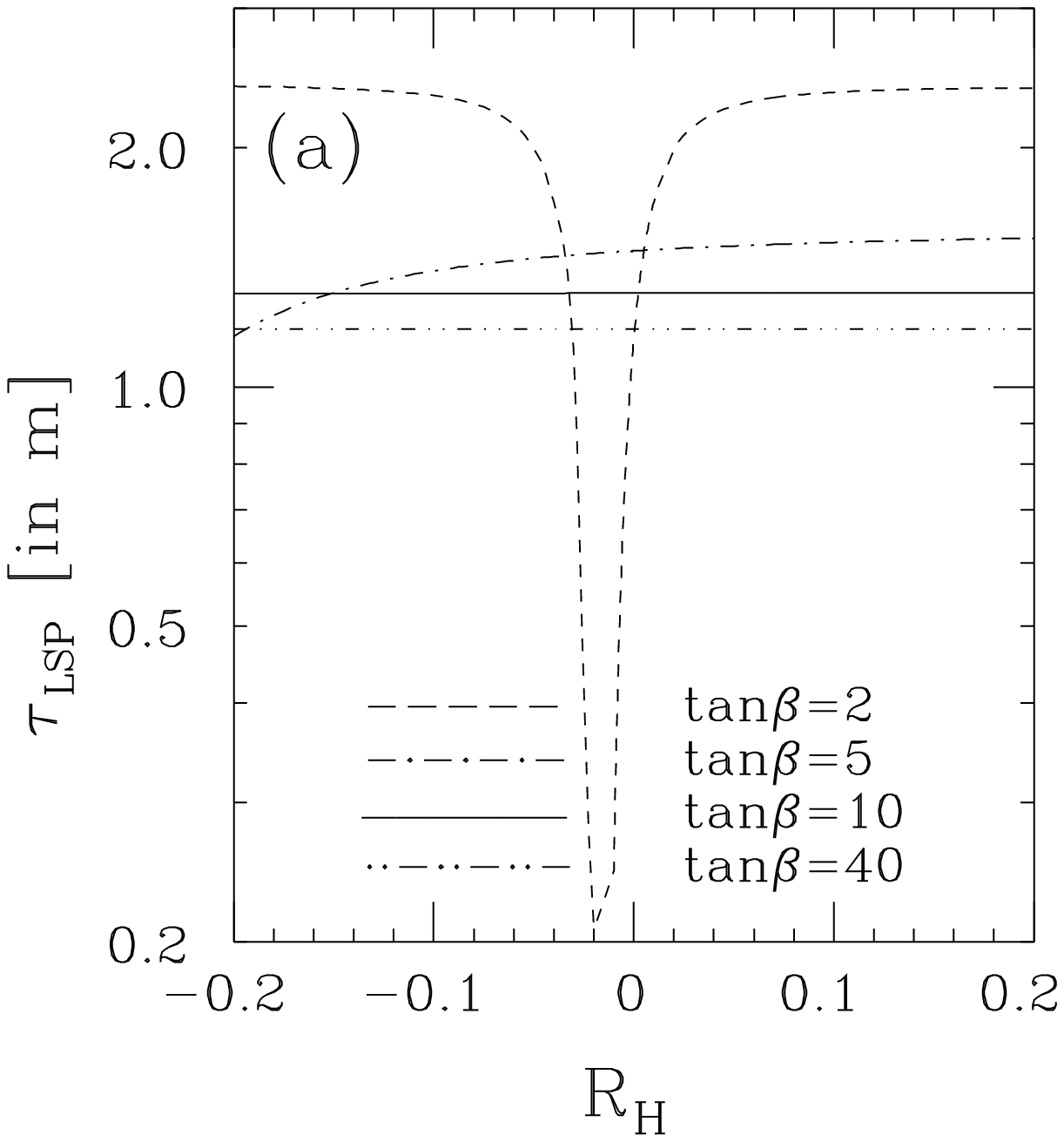}
\includegraphics{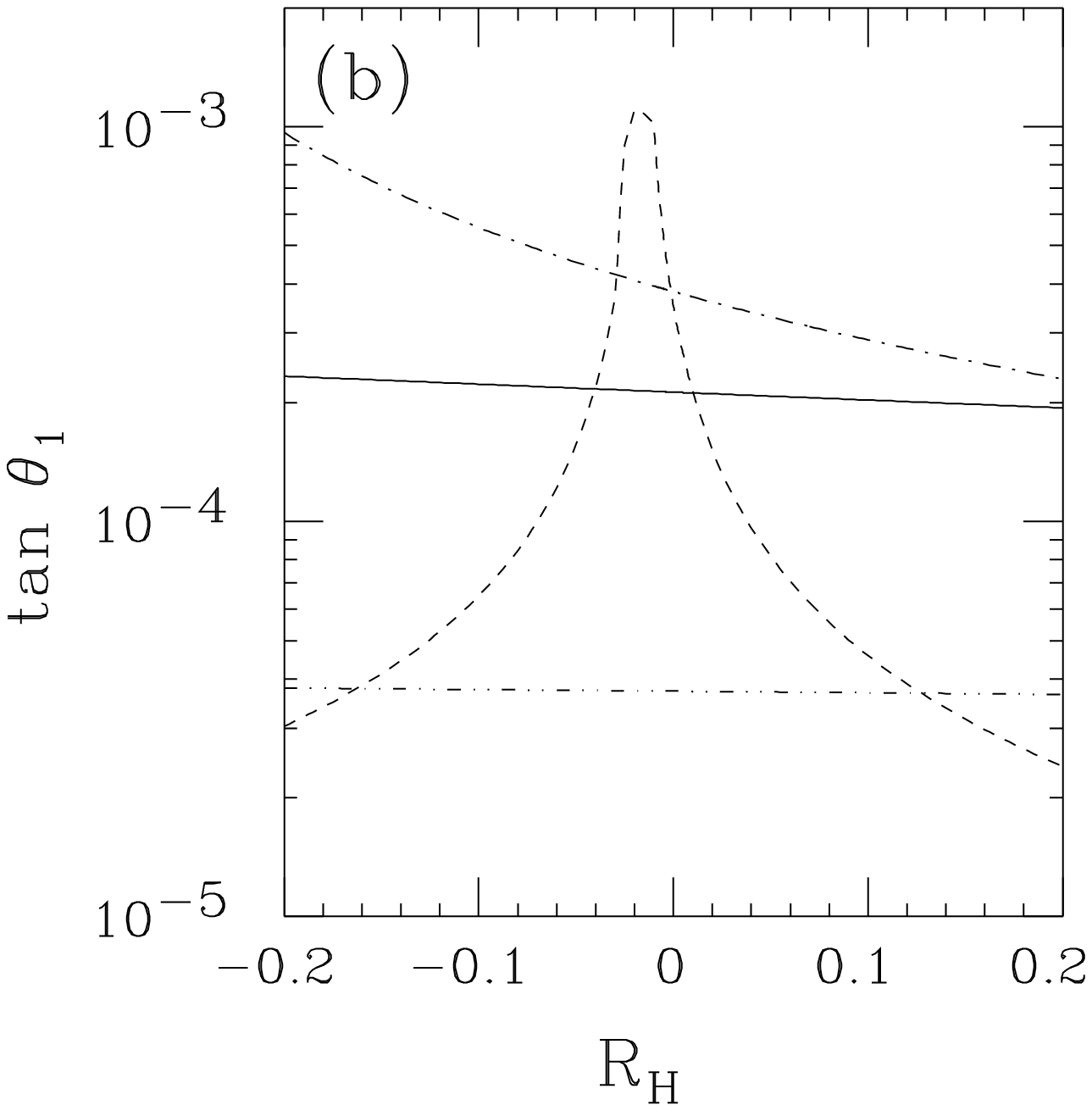}
\includegraphics{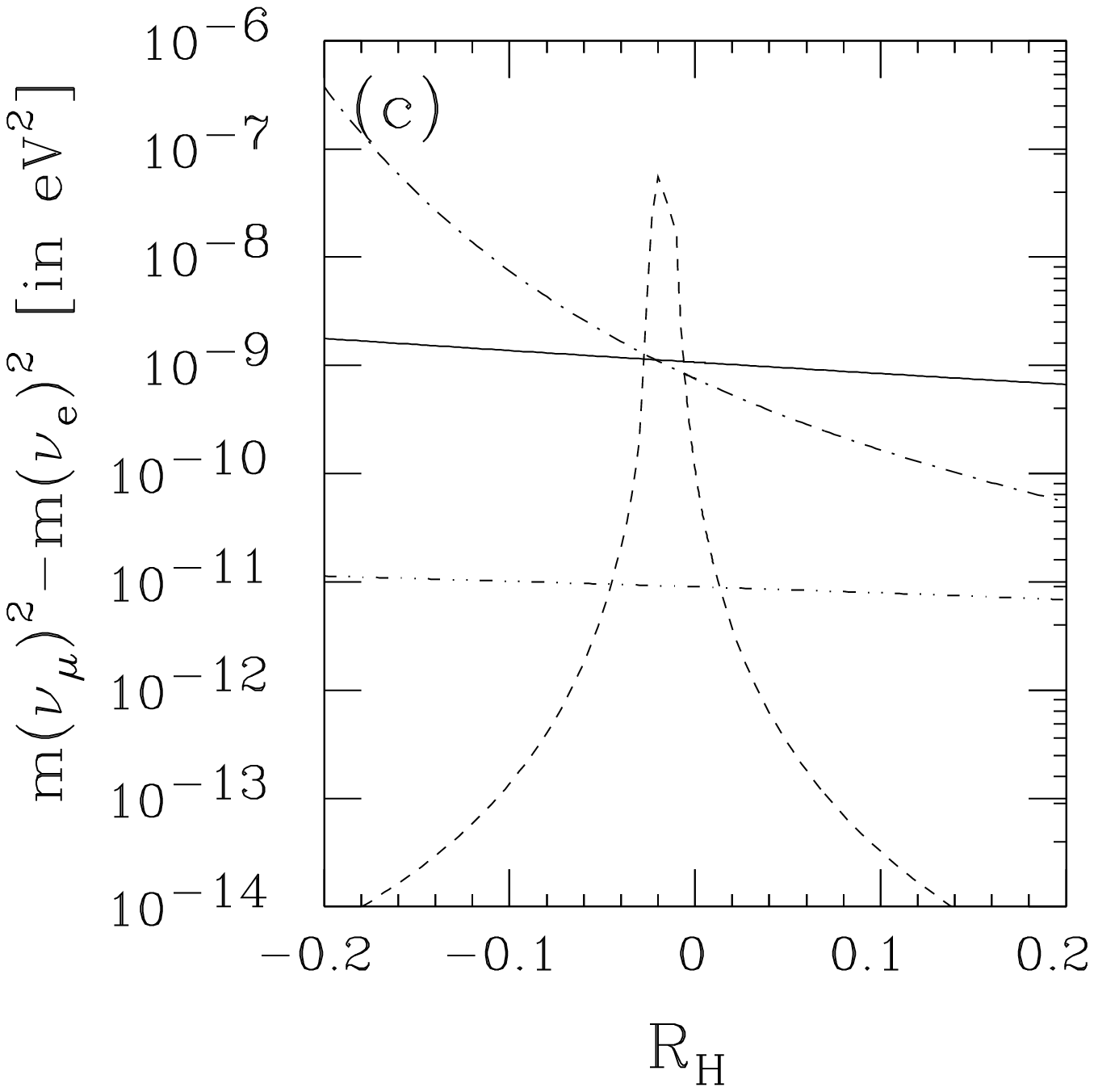}
\ffcaption{
The effects of non-universal boundary conditions
at $\mgut$ on
(a) $\tau_{LSP}$,
(b) tan$\theta_1$ and
(c) $m_{\nu_\mu}^2 - m_{\nu_e}^2$ for four
different values of $\tanb$.
We set $m_{1/2} = 120$~GeV, $\mu = 300$~GeV
and $A_0 =0$.

\bigskip
\baselineskip=15pt
{\twelverm
\baselineskip=15pt
two-body decays dominate
and differ from the full width only by a few \%\
[fig.~\ref{fig7}].
(A partial analysis of the two-body decays
has been performed previously\cite{muk}.)

\bigskip
\hskip0.25in
So far we have assumed exact universality
at $\mgut$.
However, it has been pointed out in ref.~\citenum{nir+pomerol}
that the evolution from the Planck scale $\mpl$
to $\mgut$ can already have
a significant impact on the sparticle spectrum.
This is particularly important
in SO(10) based models were
the Higgs and slepton universality is violated via gaugino
effects, since the Higgs (slepton) fields belongs to a 10-dim (16-dim)
representation
(remember: below $\mgut$ non-universal effects arise only from
Yukawa couplings while 
the non-universal effects above 
$\mgut$ arise from the gauge couplings).
We 
\phantom{suighsiuhgighusfhgiuh}}
}
\label{fig10}
\end{figure}

In fig.~\ref{fig5} we have plotted $\Gamma_{LSP}$ vs. $\tan\theta_1$.
We find that there are simple scaling relations
if $R_p$ violation is sufficiently small
(say $\tan\theta_1\lsim 0.1$):
\beqn
\Gamma_{LSP}\,, m_{\nu_x} \propto \tan^2\theta_1
\qquad(x=e,\mu,\tau)\,.
\eeqn
The parameters $\tan \theta_2$ and  $\tan \theta_3$
which govern the neutrino oscillations have only a
small impact on the LSP properties.
In fig.~\ref{fig6}
we present contours of $\tan \theta_1$
(fixed by imposing
$m_{\nu_\tau}^2 - m_{\nu_\tau}^2 = 10^{-2}~\ev^2$)
and constant $\tau_{LSP}$
in the $\tanb$--$M_{LSP}$ plane
for $\mu = \pm 2.5 m_{1/2}$.
We find that that required range of the $R_p$ violation
is $\tan \theta_1 = 10^{-x}$ ($x = 2\sim 5$)
with the upper (lower) limit corresponding
to small (large) values of $\tanb$.
The corresponding range of the LSP life-time
is $c \tau_{LSP} = 1~m \sim 0.1~mm$
(for $M_{LSP} = 40 \sim 160~\gev$)
and can easily be determined at forthcoming collider experiments.
Furthermore, we need larger $R_p$ violation (for fixed 
$m_{\nu_\tau}$) for $\mu<0$
due to cancellation among tree-level and one-loop contributions.
For $\tanb > 30\sim 40$ the one-loop contribution even dominates
over the tree-level result.
For $M_{LSP} \gsim \mz$ the

\begin{figure}
\vspace*{13pt}
\vspace*{3.6truein}      
\includegraphics{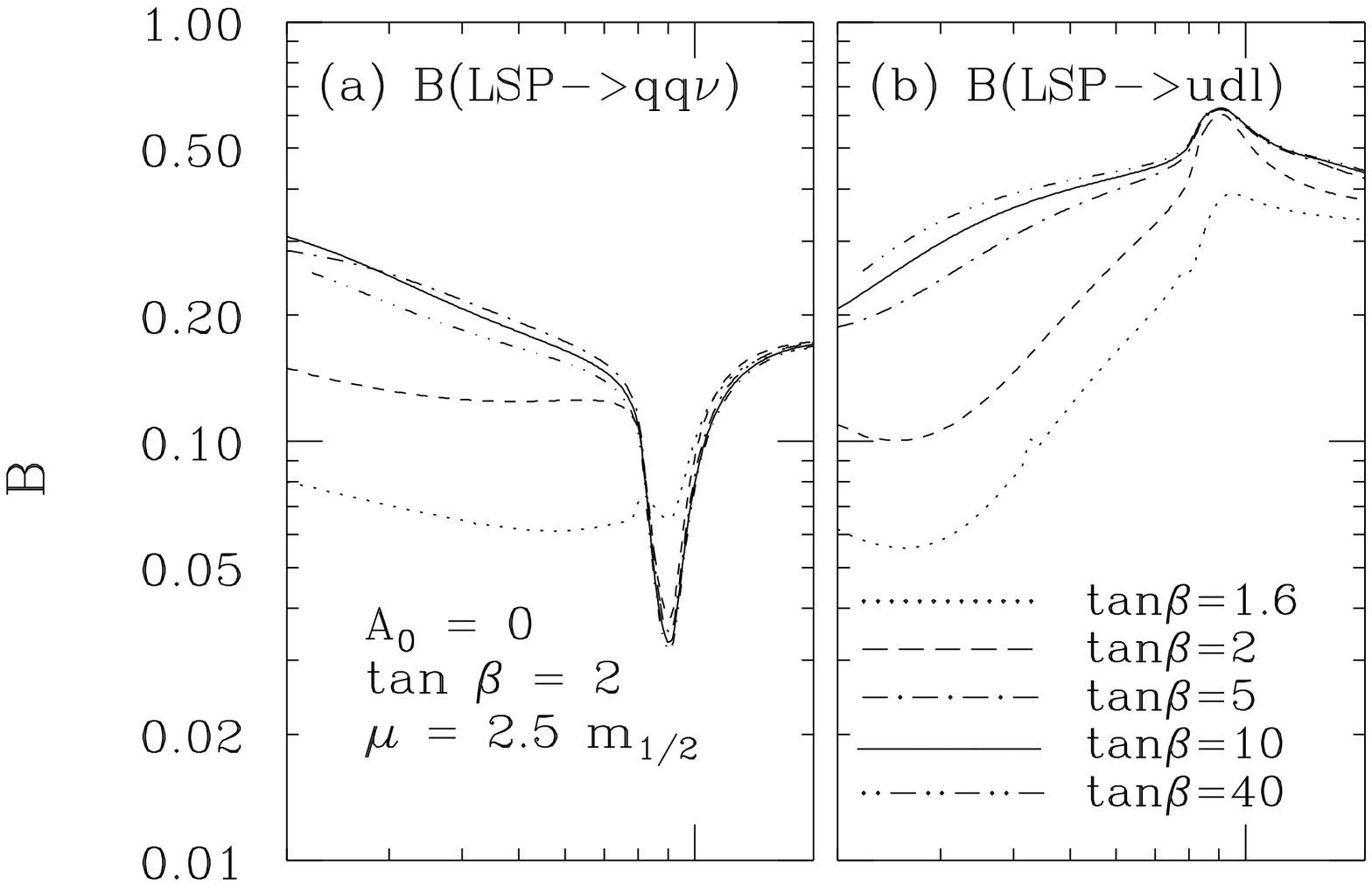}
\includegraphics{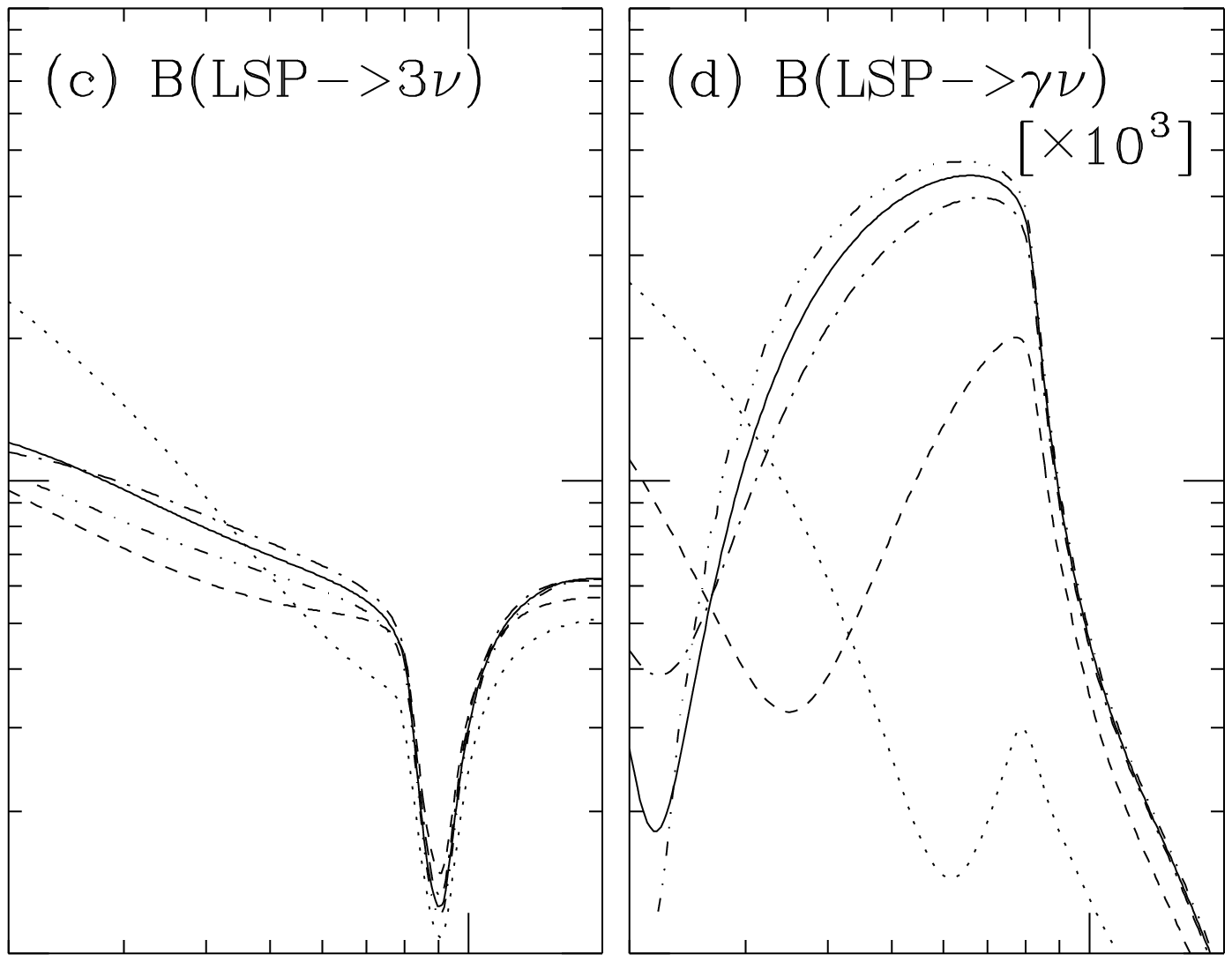}
\includegraphics{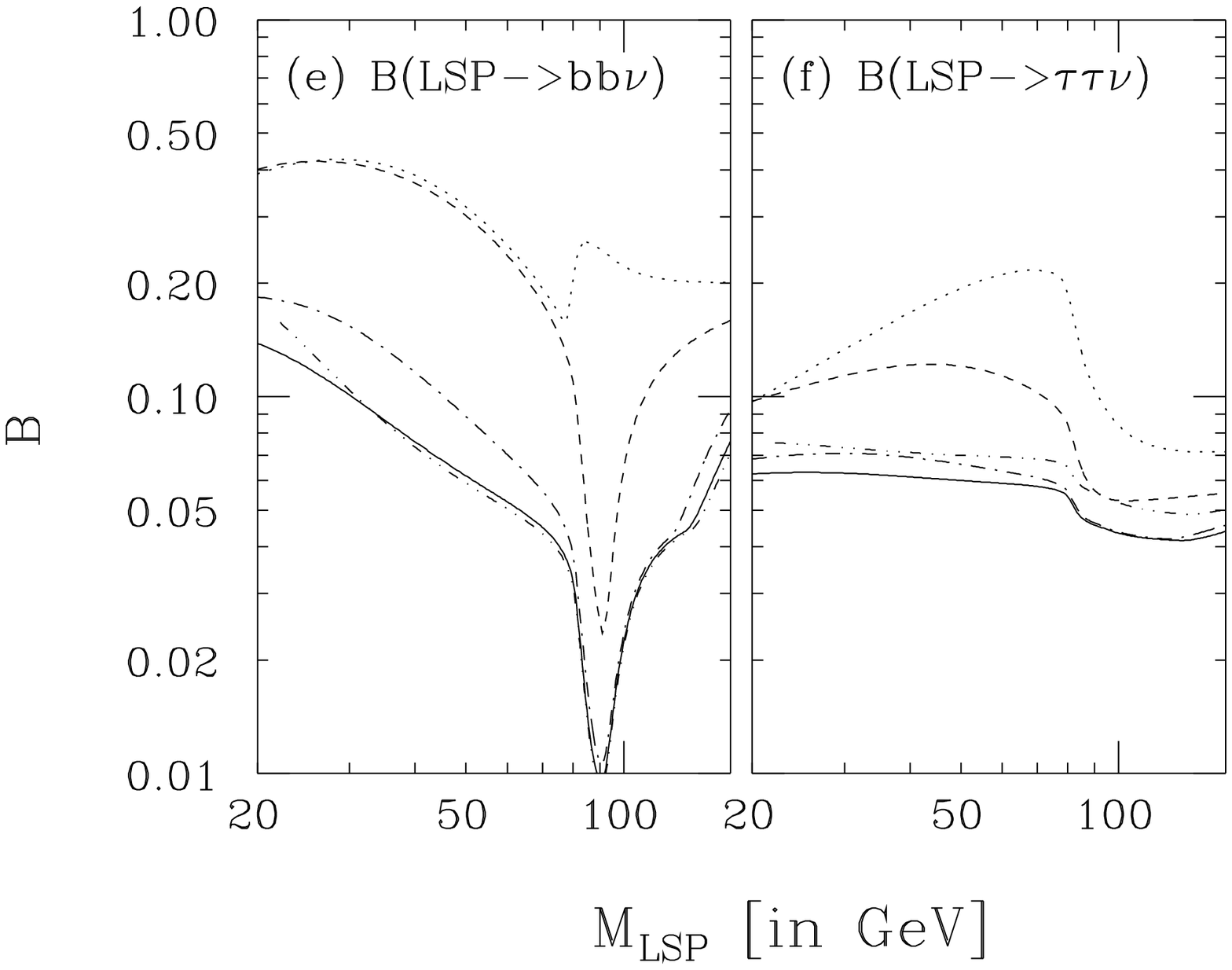}
\includegraphics{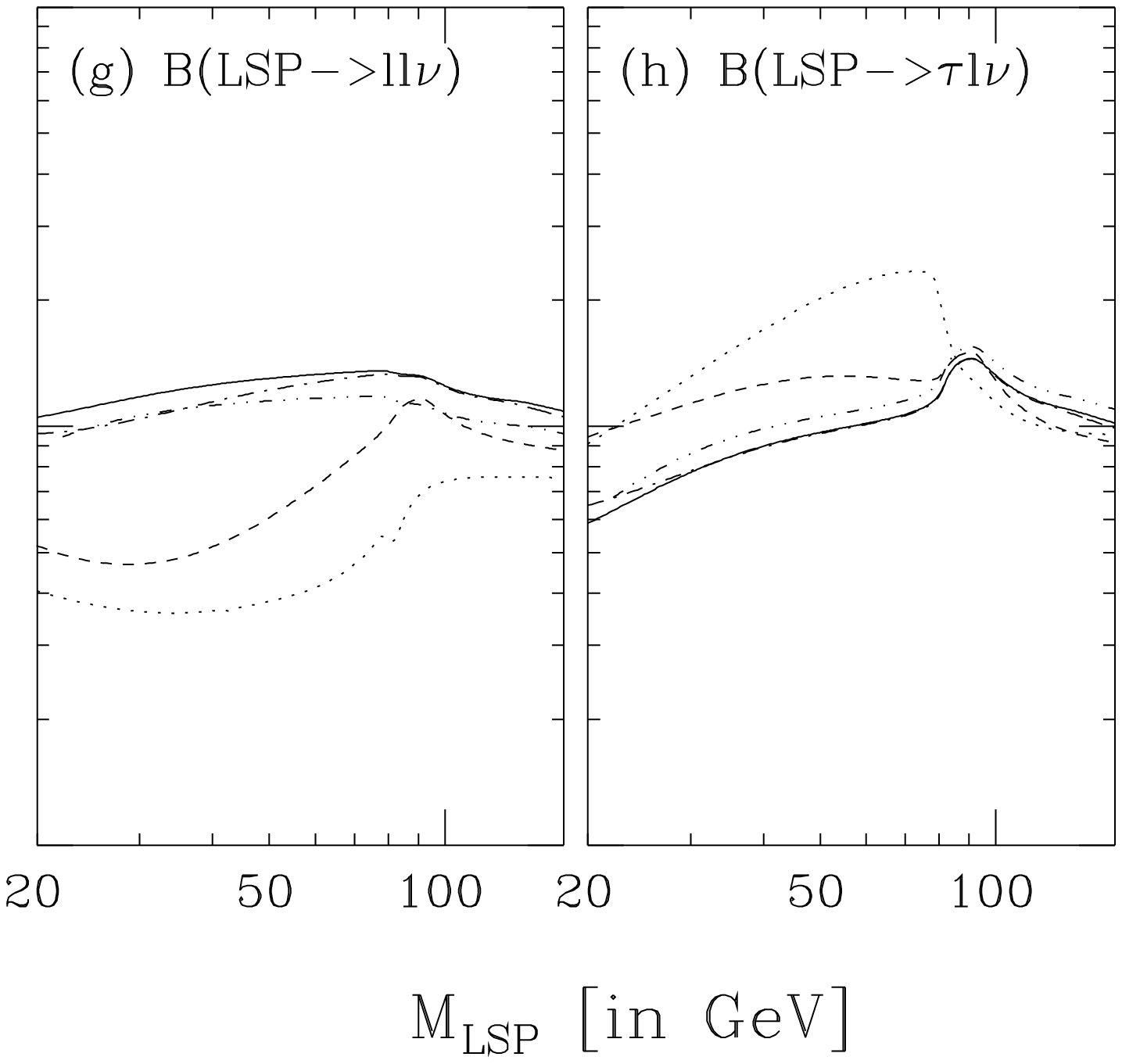}
\fcaption{Different LSP-branching fractions as functions
of $M_{LSP}$ for five values of $\tanb$.
We set $A_0=0$ and $\mu = 2.5 m_{1/2}$
}
\label{fig8}
\end{figure}

\noindent
can accomodate this effect by modifying the
boundary conditions at $\mgut$
\beqn
m_H^2(\mgut) = m_{L_0}^2(\mgut) = m_0^2 + R_H m_{1/2}^2\,,
\eeqn
where we typical expect
$R_H = 9 \alpha_{GUT}/(4\pi)\ln(\mgut/\mpl) \simeq -0.1$.

In fig.~\ref{fig10} we see that the effect of non-universal
terms is very significant (small) for small (large) values of $\tanb$ were
the down-typ Yukawa couplings are small (large).

\subsection{LSP Braching Fraction}

\begin{figure}
\vspace*{13pt}
\vspace*{2.1truein}      
\includegraphics{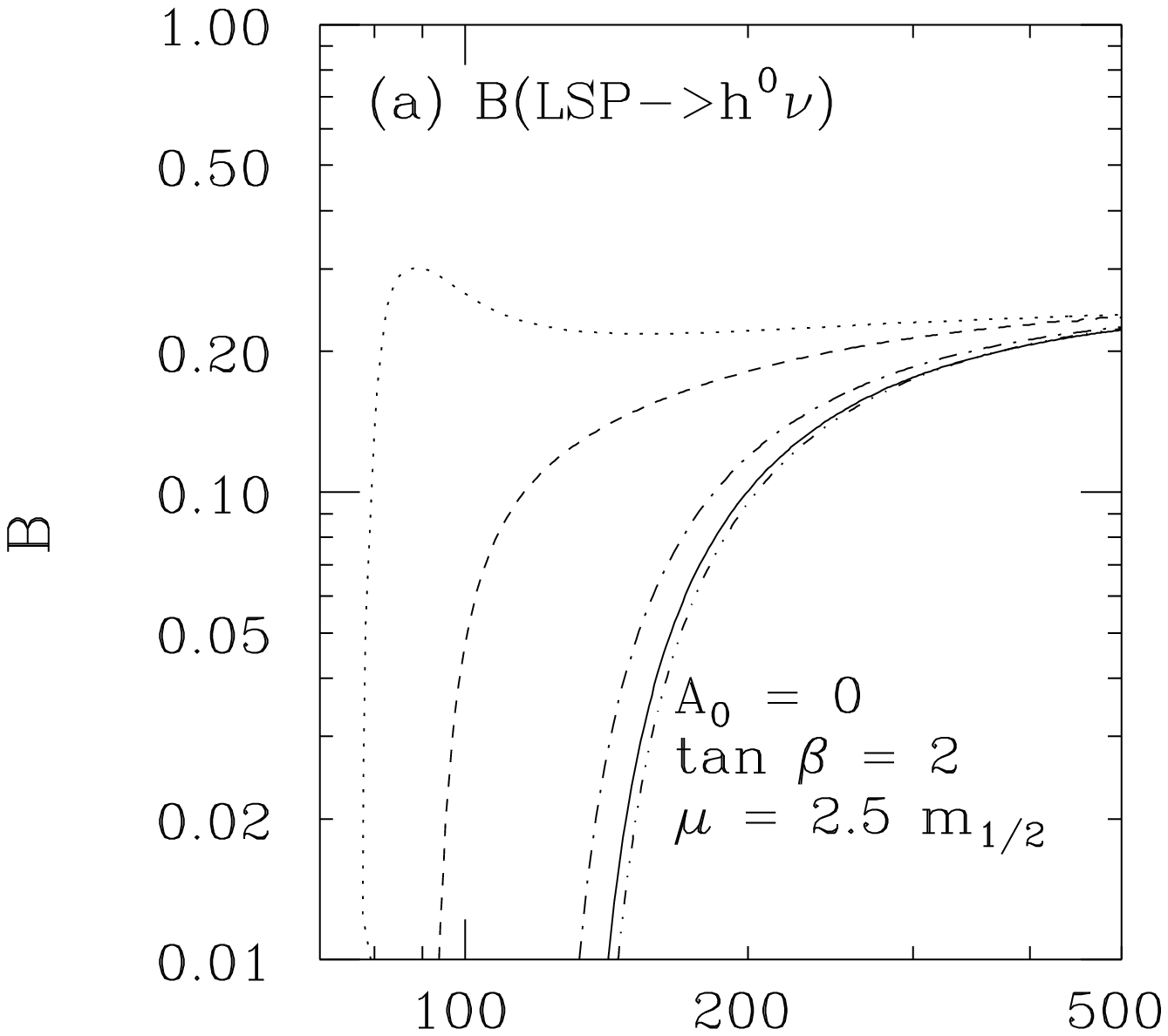}
\includegraphics{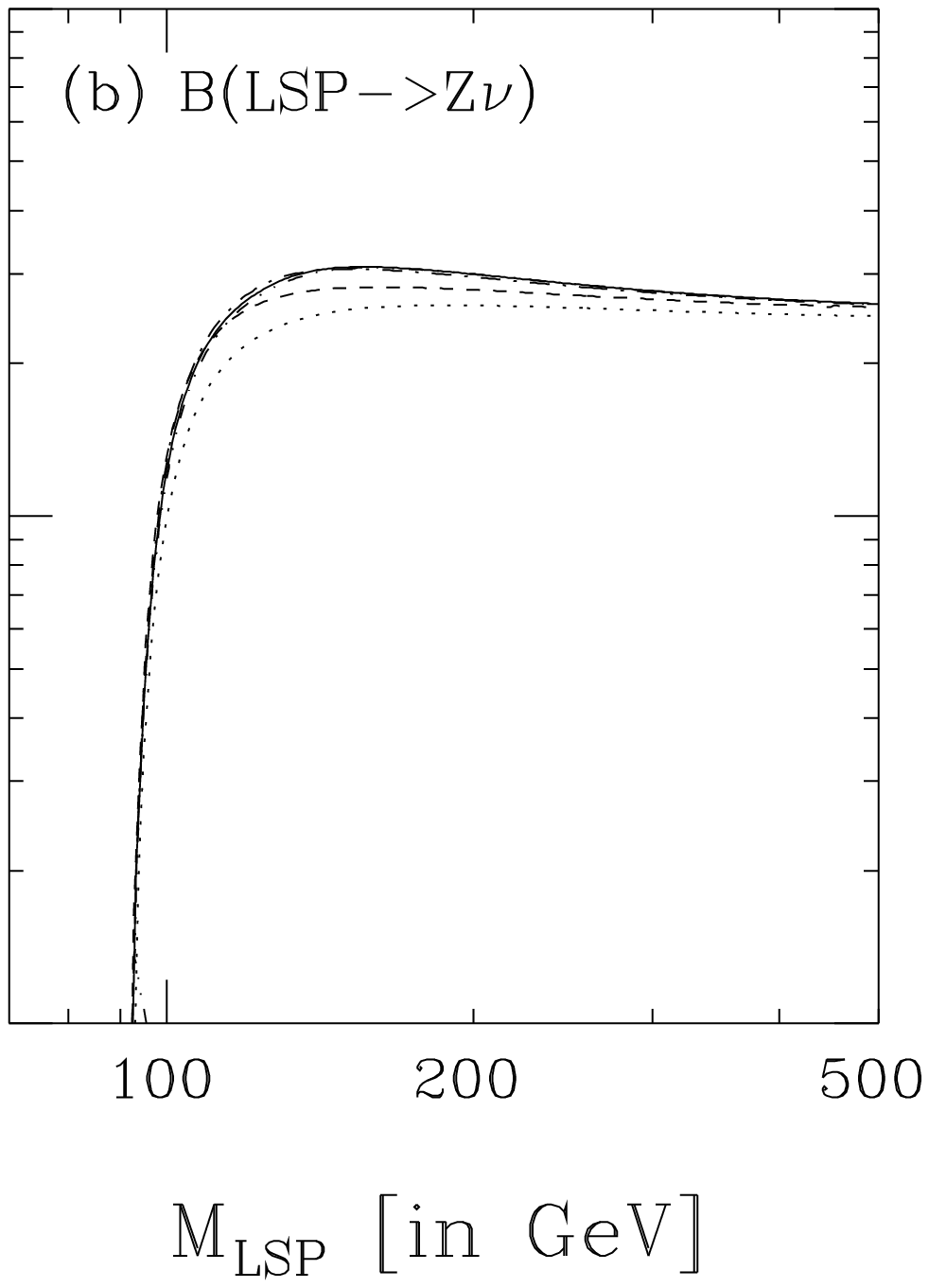}
\includegraphics{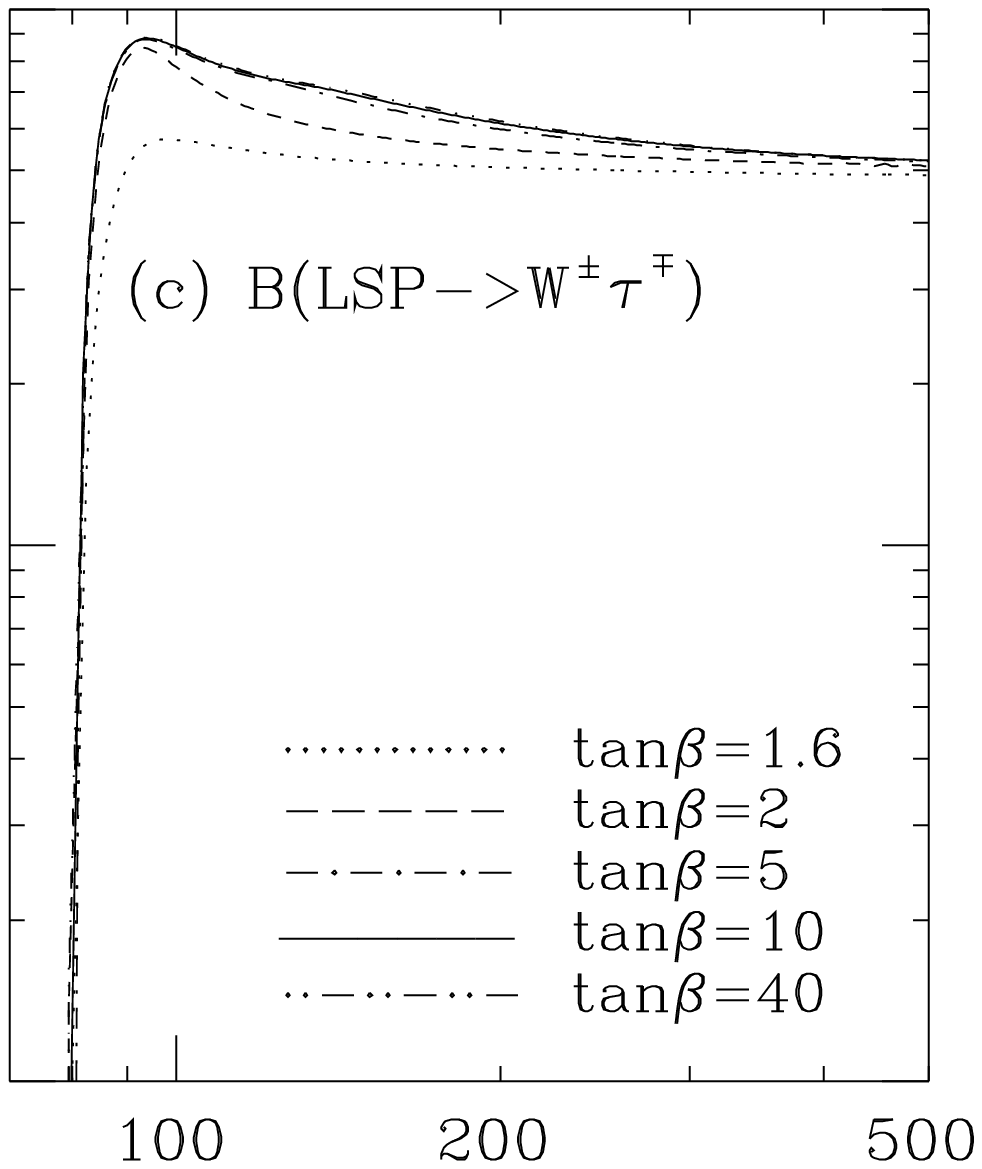}
\fcaption{
Branching fractions of different two-body decays
as functions of $M_{LSP}$ for five values of $\tanb$.
SUGRA parameters are same as in fig.~\ref{fig8}.
}
\label{fig9}
\end{figure}

So far we have used results from neutrino physics in order to
eliminate the $R_p$ breaking parameters $\tan \theta_i$ ($i= 1, 2, 3$).
However, to a good approximation this dependence drops out
if we consider the branching fractions.
In fig.~\ref{fig8} we present the branching fractions
as a function of $M_{LSP}$ for eight different channels.
The dominant decay mode is into quarks [(a) and (b) are first
two generations only; (e) is the third generation]
with a strong enhancement into $b\bar b$ (e)
for small $\tanb$. Invisible decay modes (c) are typically
below 10\%\ and the radiative decay (d) is insignificant.
The leptonic decays into $\tau^+ \tau^-$ (f),
$\ell^+ \ell^-$ ($\ell =  e, \mu)$ (g) and
$\ell^\pm \tau^\mp$ (h) is typically O(10\%).
For $M_{LSP}\gsim 100~\gev$ the situation becomes much more transparent by 
considering the two-body decays [fig.~\ref{fig9}].
Here, there are only three relevant channels with 
$B(LSP\rightarrow W^\pm \tau^\mp) \simeq 0.5$ and
$B(LSP\rightarrow Z^0   \nu),
B(LSP\rightarrow h^0   \nu) \simeq 0.25$.

\section{Conclusions}

We have investigated
the LSP phenomenology in
supersymmetric models without lepton number conservation.
In any model of this kind, lepton number is violated
spontaneously via sneutrino VEVs as well as 
explicitly. Both effects are of the same order
and have to be studied consistedly.
Assuming that Higgs-sneutrino mixing is responsible
for the observed neutrino oscillations
we find that the LSP decays inside the detector.
The life-time can be determined over a large region of the
SUSY parameter space. 
The branching fractions for
all relevant decay modes are we presented.

\section{Acknowledgements}
This work was supported in parts by the DOE under
Grants No. DE-FG03-91-ER40674 and by the
Davis Institute for High Energy Physics.

\end{document}